\documentclass[%
 reprint,
showpacs,preprintnumbers,
 amsmath,amssymb,
prb,
]{revtex4-1}

\usepackage{graphicx}
\usepackage{dcolumn}
\usepackage{bm}

\begin{document}


\title{Modeling Bloch Oscillations in Nanoscale Josephson Junctions}

\author{Heli Vora}
\email{heli.vora@nist.gov}
\author{R. L. Kautz}
\author{S. W. Nam}
\author{J. Aumentado}
\affiliation{National Institute of Standards and Technology, Boulder, Colorado, 80305}

\date{\today}

\begin{abstract}
Bloch oscillations in nanoscale Josephson junctions with a Coulomb charging energy comparable to the Josephson coupling energy are explored within the context of a model previously considered by Geigenm{\"u}ller and Sch{\"o}n that includes Zener tunneling and treats quasiparticle tunneling as an explicit shot-noise process. The dynamics of the junction quasicharge are investigated numerically using both Monte Carlo and ensemble approaches to calculate voltage--current characteristics in the presence of microwaves. We examine in detail the origin of harmonic and subharmonic Bloch steps at dc biases $I=(n/m)2ef$ induced by microwaves of frequency $f$ and consider the optimum parameters for the observation of harmonic ($m=1$) steps. We also demonstrate that the GS model allows a detailed semi-quantitative fit to experimental voltage--current characteristics previously obtained at the Chalmers University of Technology, confirming and strengthening the interpretation of the observed microwave-induced steps in terms of Bloch oscillations.
\end{abstract}


\maketitle

\section{Introduction}

Beginning in 1984, a group at Moscow State University \cite{lik84,ave85,lik85,ave86} developed a theory of nanoscale Josephson junctions for which the charging energy $E_c=e^2/(2C_j)$ of a single electron on the junction capacitance $C_j$ is comparable to or exceeds the Josephson coupling energy $E_j=\hbar I_c/(2e)$, where $I_c$ is the junction's critical current. In this limit, the charge $Q$ on the junction capacitance is replaced by its conjugate variable $\phi$ (the difference in phase between the junction electrodes) as the relevant classical variable of the system, and the dynamics of the junction are radically altered. Thus, when $E_j\gg E_c$ a dc current bias produces Josephson oscillations of frequency $f_j=2e\langle V\rangle/h$ (where $\langle V\rangle$ is the average junction voltage),\cite{jos62} whereas when $E_c\gtrsim E_j$ the Moscow State theory predicts that a dc bias will produce Bloch oscillations of frequency $f_b=\langle I_p\rangle/(2e)$, where $\langle I_p\rangle$ is the average pair current through the junction. Such Bloch oscillations in nanoscale junctions were verified experimentally beginning in 1991 by a group at Chalmers University of Technology \cite{kuz91,hav91,kuz92a,kuz92b,kuz92c,kuz93,kuz94a,kuz94b} with the observation of a microwave-induced peak in the $d\langle V\rangle/dI$ curve at a bias current $I=2ef$, where $f$ is the microwave frequency. This peak demonstrates that Bloch oscillations can phase lock with applied microwaves and suggests, as noted by the Moscow State group, that a nanoscale junction might be used to make a quantum standard for current, just as larger junctions are used to make quantum voltage standards. \cite{ham00} The question that remains is whether the width of the peak in $d\langle V\rangle/dI$ can be narrow enough to create a standard of metrological precision. In attempting to answer this question, we adopt a model of nanoscale junctions explored by Geigenm{\"u}ller and Sch{\"o}n (GS) that explicitly incorporates the shot noise of quasiparticle (single-electron) tunneling.\cite{gei88} We begin with an introduction to junction dynamics and derive the equations of motion of a nanoscale Josephson junction. We then explain each calculation approach in detail, namely the Monte Carlo and the ensemble calculations. Using the advantage of faster computational speed in the ensemble approach, our model offers insight for the parameters needed to obtain current steps of metrological interest. Ultimately, we show that this model provides a semi-quantitative explanation of the Chalmers experiments and demonstrates the possibility of creating a metrologically precise current standard if the errors from quasiparticle tunneling can be reduced. 

\section{Junction Dynamics}

\begin{figure}[b]
\includegraphics{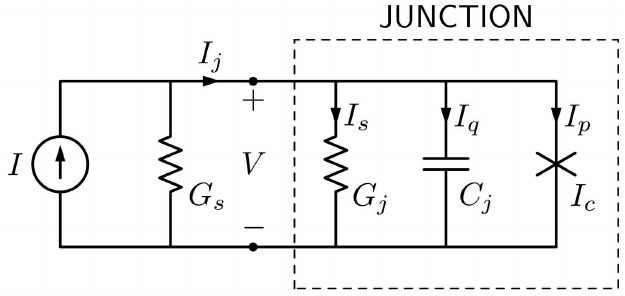}
\caption{\label{fig:circuit} Circuit diagram of a Josephson junction driven by a current source.}
\end{figure}

The basic circuit considered here, shown in Fig.~\ref{fig:circuit}, consists of a superconducting tunnel junction driven by a current source $I$ with source conductance $G_s=1/R_s$. The junction itself comprises a single-electron tunneling element of conductance $G_j=1/R_j$, a capacitor $C_j$, and a Josephson element associated with pair tunneling characterized by a critical current $I_c$. As a two-terminal device, the junction is entirely defined by the relation between the current $I_j$ and voltage $V$ at its terminals. However, given that the tunneling elements are highly nonlinear and energy can be stored in both the capacitor and the Josephson element, this relation is generally complex and depends critically on whether $E_j\gg E_c$ or $E_c\gtrsim E_j$. \cite{ave86,gei88}

It is necessary to differentiate between the source conductance $G_s$, which can support a continous current flow, and the junction conductance $G_j$, which represents stochastic quasiparticle tunneling. The source conductance represents a Norton equivalent of a series isolation resistor $R_s=1/G_s$ that must be much larger than the resistance quantum $R_Q=R_K/4=h/4e^2$ to obtain the quantum effects that lead to Bloch oscillations. \cite{lik85} The following analyses also require that the thermal energy $kT_j$ of the junction electrons be much less than $E_j$ for large-area junctions and less than both $E_j$ and $E_c$ for nanoscale junctions.

For junctions with $E_j\gg E_c$, quantum calculations simplify to yield the following relations between $I_j$  and $V$,\cite{mcc68,ste68}
\begin{eqnarray}
I_j&=&G_jV+C_j\frac{dV}{dt}+I_c\sin{\phi}\label{eq:Ij},
\\
\frac{d\phi}{dt}&=&\frac{2e}{\hbar}V\label{eq:dphi},
\end{eqnarray}
 with the junction phase $\phi$ acting as a classical intermediary. Although nonlinear, these equations are relatively simple and valid for frequencies much less than the energy-gap frequency $f_g=(\Delta_a+\Delta_b)/h$ (where $\Delta_a$ and $\Delta_b$ are the superconducting energy gaps of the junction electrodes). They suffice to describe the dynamics of large-area junctions over a wide range of conditions.

For junctions with $E_c\gtrsim E_j$, the voltage--current relation cannot be realistically modeled so simply. In this case there is an interplay between the capacitor and the Josephson element that must be handled quantum mechanically.\cite{and64} To develop this idea, we consider these two elements as an isolated system.

Classically, the capacitor's energy is a function of the charge, $Q^2/(2C_j)$, while that of the Josephson element is a function of the phase, $\int I_pVdt=\int E_j\sin\phi\, d\phi=-E_j\cos\phi$. Quantum mechanically, the Hamiltonian of  the system is,
\begin{equation}
H=\frac{Q^2_{op}}{2C_j}-E_j\cos\phi\;,
\end{equation}
where the charge operator $Q_{op}$ is conjugate to the phase $\phi$ and takes the form $Q_{op}=(2e/i)\partial/\partial\phi$. The Hamiltonian thus becomes,
\begin{equation}
H=-4E_c\frac{\partial^2}{\partial\phi^2}-E_j\cos\phi\;,
\end{equation}
which is an exact analog to the Hamiltonian for a particle in a sinusoidal potential, which in turn is a one-dimensional model for conduction electrons in a crystalline solid.\cite{sla52} Thus, by solving the eigenvalue problem $H\psi=E\psi$, we will find stationary quantum states of our tunnel junction analogous to the Bloch states of a 1-D crystal.

In the present case, the Bloch states take the form,
\begin{equation}
\psi_{\tilde Q}(\phi)=P_{\tilde Q}(\phi)\, e^{i\phi\tilde Q/2e}\,,
\end{equation}
where $P_{\tilde Q}(\phi)$ is a periodic function, $P_{\tilde Q}(\phi+2\pi)=P_{\tilde Q}(\phi)$, and $\tilde Q$ is an index of the eigenstate called the quasicharge analogous to the quasimomentum of electrons in crystals. By construction, $\psi_{\tilde Q}(\phi)$ is a state of definite quasicharge but indefinite phase. Expanding $P_{\tilde Q}(\phi)$ in a Fourier series allows numerical evaluation of the eigenstates and the corresponding eigenenergies $E(\tilde Q)$.

\begin{figure}[b]
\includegraphics{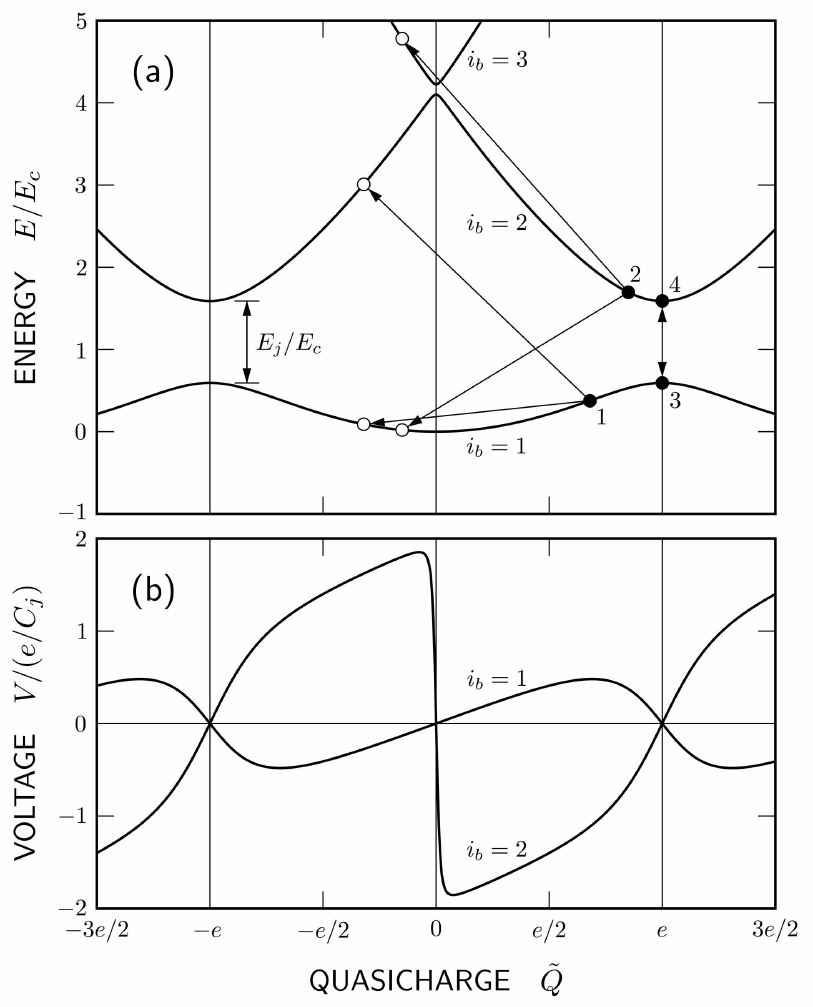}
\caption{\label{fig:energy} Energy (a) and voltage (b) as a function of quasicharge for a nanoscale junction with $\varepsilon_j=1$. Directed lines indicate single-electron tunneling processes originating at points 1 and 2 and Zener tunneling originating at points 3 and 4.}
\end{figure}

The calculated eigenenergies in units of $E_c$ depend only on the ratio $\varepsilon_j=E_j/E_c$ and are shown in Fig.~\ref{fig:energy}(a) for $\varepsilon_j=1$. As seen here, the original energy parabola $Q^2/(2C_j)$ of the capacitor is split into a series of energy bands with a gap of $E_j$ between the first and second bands and smaller gaps between higher bands. The eigenstates $\psi_{i_b}(\tilde Q)$ and energies $E_{i_b}(\tilde Q)$ are specified by a band index $i_b$ and the quasicharge $\tilde Q$. Because $E$ is $2e$ periodic in $\tilde Q$, it is possible to restrict attention to the first Brillouin zone,  $-e\le\tilde Q\le e$, although the extended zone scheme is often useful.

The band structure of Fig.~\ref{fig:energy}(a) is key to understanding the dynamics of nanoscale junctions. As long as the external forcing does not change too rapidly, the Josephson element and capacitor taken together will be found in an eigenstate with a definite band index $i_b$, quasicharge $\tilde Q$, and energy $E_{i_b}(\tilde Q)$. Thus, $i_b$ and $\tilde Q$ are the classical state variables of the combined Josephson-capacitor circuit element. The voltage of this element, which sets the voltage $V$ of the entire circuit,  is simply the derivative of $E$ with respect to $\tilde Q$ (just as the voltage of a capacitor is $d(Q^2/2C)/dQ$),
\begin{equation}
V=V_{i_b}(\tilde Q)=\frac{dE_{i_b}(\tilde Q)}{d\tilde Q}\label{eq:V}.
\end{equation}
 Thus, the voltage is also fixed by the state variables $i_b$ and $\tilde Q$ and is periodic in $\tilde Q$. Examples of the functions $V_{i_b}(\tilde Q)$ are plotted in Fig.~\ref{fig:energy}(b).

\subsection{Bloch Oscillations}

Changes in $i_b$ and $\tilde Q$ derive from three sources: the current $I_j$, single-electron tunneling associated with $G_j$, and a process known as Zener tunneling. Both tunneling processes change $i_b$ and $\tilde Q$ instantaneously and, perhaps surprisingly, do not contribute directly to the current flowing through the junction. We discuss these processes in the following sections.

The effect of $I_j$ is analogous to that of an electric field acting on an electron in a crystalline solid, which changes the electron's quasimomentum without changing the band index. The equivalent change in the quasicharge for the Josephson system is simply,
\begin{equation}
\frac{d\tilde Q}{dt}=I_j\label{eq:dQ}.
\end{equation}
Thus, a steady current $I_j$ causes the quasicharge $\tilde Q$ to increase uniformly in time, as might be expected for an ordinary capacitor. However, considering the implied advance of $\tilde Q$ through the energy and voltage bands of  Fig.~\ref{fig:energy} for say $i_b=1$, we conclude that both $E$ and $V$ will oscillate regularly with a period of $2e/(d\tilde Q /dt)=2e/I_j$. These are the expected Bloch oscillations.

Equations (\ref{eq:V}) and (\ref{eq:dQ}) define the relation between $I_j$ and $V$ for nanoscale junctions in the limit of slow motion,  $I_j\ll e/(R_KC_j)$, when single-electron and Zener tunneling are neglected. These equations can be viewed as the dual of Eqs.~(\ref{eq:Ij}) and (\ref{eq:dphi}) in that Eq.~(\ref{eq:V}) expresses $V$ as a periodic function of the internal state variable $\tilde Q$ while Eq.~(\ref{eq:Ij}) expresses $I_j$ as a periodic function of the internal state variable $\phi$, and Eq.~(\ref{eq:dQ}) equates $d\tilde Q/dt$ to $I_j$ while Eq.~(\ref{eq:dphi}) relates $d\phi/dt$ to $V$. That is, the relationships are exactly similar with the roles of $I_j$ and $V$ reversed. Thus, it makes sense that nanoscale junctions might lead to quantized currents just as large-area junctions yield quantized voltages. However, while there are no known corrections to Eq.~(\ref{eq:dphi}), single-electron tunneling leads to sudden shifts in $\tilde Q$, not included in Eq.~(\ref{eq:dQ}), that compromise the precision of the equivalent current standard.

We might envision a Bloch oscillation in the first band as beginning with the high-energy state at $\tilde Q=-e$ and proceeding until the applied current $I_j>0$ raises the quasicharge to $\tilde Q=e$, where it reaches another high energy state. This picture would make sense if the junction were simply a capacitor, except that $V=0$ at both the beginning and end of the process, and the process will exactly repeat itself  as $I_j$  forces more charge through the junction. At $\tilde Q=e$, we can choose to say that the system undergoes a Bloch reflection to the equivalent quasicharge $\tilde Q=-e$, suggesting that pair tunneling discharges the junction at this point. However, no special event actually occurs at $\tilde Q=e$, and we could as easily choose to say that the quasicharge continues to increase beyond $\tilde Q=e$. Nevertheless, it is clear that pair tunneling occurs with certainty each time $\tilde Q$ increases by $2e$.  This process is known as coherent tunneling to distinguish it from the sudden, randomly timed character of common tunneling events. The certainty of coherent pair tunneling makes it especially attractive as the basis for a quantum current standard.

\subsection{Single-Electron Tunneling}

In large-area junctions, the tunneling of a single electron changes the energy stored on $C_j$ by an amount insignificant compared to $E_j$, and single-electron tunneling is simply represented by a continuous normal current flowing through $G_j$. In nanoscale junctions, by contrast, the tunneling of a single electron changes the state variable $\tilde Q$ by $e$ and possibly the band index $i_b$ by $\pm1$,  which typically produces a significant, instantaneous change in the junction's energy, $E_{i_b}(\tilde Q)$. Thus, the tunneling of one electron completely disrupts the otherwise continuous Bloch oscillations.

Single-electron tunneling is a Poisson process with an instantaneous rate given by, \cite{gei88}
\begin{equation}
\Gamma=\frac{G_j\Delta E/e^2}{\exp(\Delta E/kT_j)-1}\label{eq:Gamma},
\end{equation}
where $\Delta E$ is the difference in energy between the initial and final states,
\begin{equation}
\Delta E=E_{i_b'}(\tilde Q')-E_{i_b}(\tilde Q)\label{eq:DeltaE1}.
\end{equation}
Here the final quasicharge can be taken as either $\tilde Q'=\tilde Q +e$ or $\tilde Q-e$ as these states differ by $2e$ and are equivalent. On the other hand, the final band index $i_b'$ is restricted to being either 1 or 2 if $i_b=1$ (as indicated for initial state 1 in Fig.~\ref{fig:energy}(a)) or $i_b'=i_b\pm1$ if $i_b\ge2$ (as indicated for initial state 2 in Fig.~\ref{fig:energy}(a)).\cite{gei88}

Because single-electron tunneling shifts $\tilde Q$ by $e$, it interrupts the ongoing Bloch oscillation, disrupting its periodicity and compromising the accuracy of the proposed current standard. Minimizing the tunneling rate $\Gamma$ by lowering the junction temperature is advantageous in this regard both because it helps eliminate thermally activated tunneling and because it lowers $G_j$ by freezing out unpaired electrons. Indeed, theory suggests that sufficiently low temperatures would virtually eliminate single-electron tunneling.\cite{har74} There may be limits to this stratagem, however, as we discuss in section VII.

\subsection{Zener Tunneling}

 The final element to be considered in the dynamics of nanoscale junctions is Zener tunneling, a process in which the junction state changes abruptly without the transport of charge. In particular, Zener tunneling occurs when the system jumps from the energy maximum of one band to the energy minimum of the next higher band or vice versa as it passes through the maximum or minimum point. As indicated in Fig.~\ref{fig:energy}(a), this might result in an upward leap from point 3 to point 4 or a downward leap from point 4 to point 3.

 The probability of Zener tunneling from band $i_b$ to band $i_b+1$ or vice versa at the point of minimum separation between the bands is \cite{gei88},
 \begin{equation}
 P_Z=\exp\left[-\frac{\pi e(\Delta E)^2}{4\hbar E_ci_b|I_j|}\right]\label{eq:PZ},
 \end{equation}
 where $\Delta E$ is the difference in energy between the initial and final states,
 \begin{equation}
 \Delta E=E_{i_b+1}(\tilde Q)-E_{i_b}(\tilde Q)\label{eq:DeltaE2},
 \end{equation}
 and $\tilde Q=e$ or $0$, depending on whether $i_b$ is odd or even.

 Even though Zener tunneling does not directly interfere with Bloch oscillations, the higher energy of the upper bands increases the probability of single-electron tunneling. Considering the rate of single-electron tunneling from band 2 to band 1 ($\Delta E<0$) in the limit of low temperatures, we have $\Gamma =-G_j\Delta E/e^2$, making such events generally more likely than those within the first band, where $|\Delta E|$ is smaller. Thus Bloch oscillations will be interrupted less often if the gap $E_j$ between the first and second bands is as large as possible, minimizing $P_Z$ and keeping Bloch oscillations within the lower band. On the other hand, if $E_j$ is much larger than $E_c$, the junction will not obey the rules of nanoscale junctions and Bloch oscillations will disappear. While the optimum compromise between large and small $E_j$ is unknown, it is likely to occur for $E_j\simeq E_c$ or $\varepsilon_j\simeq 1$.

 \section{Monte Carlo Simulation}

 The dynamical behavior of a nanoscale junction is specified by Eqs.~(\ref{eq:V})--(\ref{eq:DeltaE2}). When these are combined with equations for the current source,
\begin{eqnarray}
I&=&I_0+I_1\sin(2\pi ft),
\\
 &=&I_j+G_sV_{i_b}(\tilde Q),
 \end{eqnarray}
 which includes a dc bias $I_0$ and a microwave bias of amplitude $I_1$ and frequency $f$, we obtain the complete circuit model to be considered here. Our goal is to calculate the average voltage $\langle V\rangle$ and its derivative $d\langle V\rangle/dI_0$ as a function of $I_0$. However, the random nature of single-electron and Zener tunneling imply that the differential equation relating $V$ and $I$ is stochastic, in contrast to the deterministic Eqs.~(\ref{eq:Ij}) and (\ref{eq:dphi}) for large-area junctions.

 The most direct approach to computing the average voltage in nanoscale junctions\cite{gei88} is simply to follow the state $(i_b, \tilde Q)$ of the junction over a long period of time as it is driven by $I_j$ according to Eq.~(\ref{eq:dQ}) and experiences sudden random changes according to the probabilities specified by Eqs.~(\ref{eq:Gamma}) and (\ref{eq:PZ}). Such a Monte Carlo simulation is relatively easy to program, but an accurate evaluation of $\langle V\rangle$ requires tracking the system for a large number of drive cycles, and the evaluation of $d\langle V\rangle/dI_0$ by taking numerical differences is problematic. Nonetheless, the Monte Carlo approach provides valuable insight into the behavior of nanoscale junctions.

Before considering a specific example, it is useful to rewrite the equations of motion in terms of  dimensionless variables and parameters. If we generically adopt dimensionless variables for current $i=I/(e/R_jC_j)$, voltage $v=V/(e/C_j)$, energy $\varepsilon=E/E_c$, quasicharge $q=\tilde Q/e$, tunneling rate $\gamma=\Gamma R_jC_j$, and time $\tau=t/(R_jC_j)$, then the equations of motion become,
\begin{eqnarray}
v&=&v_{i_b}(q)=\frac{1}{2}\frac{d\varepsilon_{i_b}(q)}{dq}\label{eq:v},
\\
\frac{dq}{d\tau}&=&i_j=i_0+i_1\sin(\omega\tau)-g_sv_{i_b}(q)\label{eq:dq},
\\
\gamma&=&\frac{\Delta\varepsilon/2}{\exp(\Delta\varepsilon/t_j)-1}\label{eq:gamma},
\\
P_Z&=&\exp\left[-\frac{(\Delta\varepsilon)^2}{4\alpha i_b|i_j|}\right]\label{eq:PZ1},
\end{eqnarray}
where it is understood that $\Delta\varepsilon=\varepsilon_{i_b'}(q')-\varepsilon_{i_b}(q)$ in Eqs.~(\ref{eq:gamma}) and (\ref{eq:PZ1}) is the energy difference between the final and initial states appropriate to single-electron and Zener tunneling, respectively. The system modeled by Eqs.~(\ref{eq:v})--(\ref{eq:PZ1}) is specified by seven dimensionless parameters:
\begin{eqnarray}
\varepsilon_j&=&E_j/E_c,\label{eq:ej}
\\
g_s&=&G_s/G_j,
\\
t_j&=&k_BT_j/E_c,
\\
i_0&=&I_0/(e/R_jC_j),
\\
i_1&=&I_1/(e/R_jC_j),
\\
\omega&=&2\pi fR_jC_j.\label{eq:omega}
\end{eqnarray}
Throughout the remainder of this paper, we explore the voltage--current characteristics of nanoscale junctions within this parameter space. Note that even though $\alpha$ in GS is defined in relation to $R_j$, \cite{gei88} in our model we treat it as a free parameter that controls the Zener tunneling strength. It is set to zero ($\alpha=0$) whenever Zener tunneling needs to be omitted. In the current section, we demonstrate that by using the Monte Carlo technique for a simple case, with $t_j=\alpha=0$, we are able to reproduce results from GS Fig.~8.\cite{gei88} With these restrictions, motion is confined to the first energy band, as there is no thermal energy to allow single-electron tunneling to higher bands and Zener tunneling is prohibited.

\subsection{dc Bias}

\begin{figure}[t]
\includegraphics{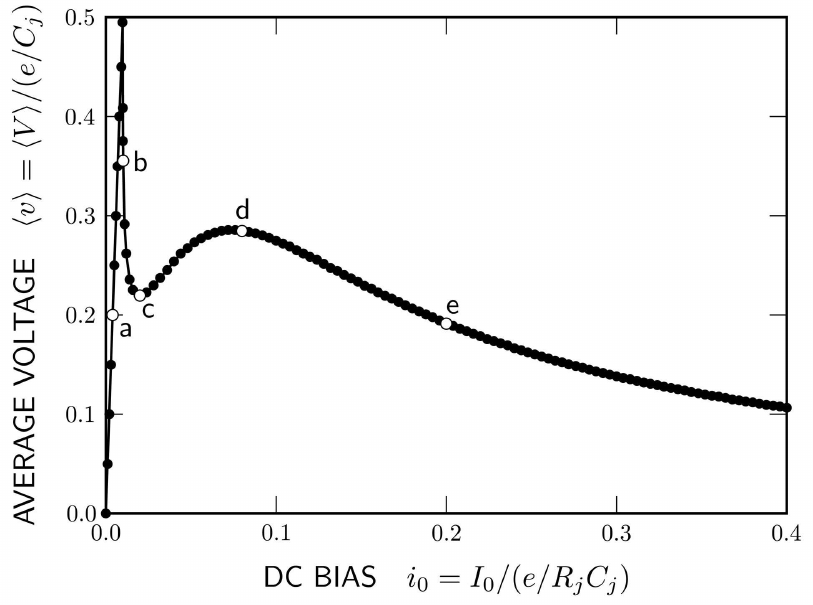}
\caption{\label{fig:dciv1} Average voltage as a function of dc bias in the presence of single-electron tunneling and in the absence of Zener tunneling for $\varepsilon_j=0.2$, $t_j=0$, $\alpha=0$, $g_s=0.02$, and $i_1=0$, computed by Monte Carlo simulation with an averaging time of $4\times10^6R_jC_j$. Parameters are chosen to match Fig. 8 from GS.\cite{gei88}}
\end{figure}

We begin with the simplest case: the  $\langle v\rangle$--$i_0$ curve for zero temperature, no Zener tunneling, and dc bias only. Choosing $\varepsilon_j=0.2$ and $g_s=0.02$ as the only non-zero parameters, we obtain the result shown in Fig.~\ref{fig:dciv1}. This curve is almost identical to that in Fig.~8 of GS at dc biases above about $i_0=0.05$, but the sharp resonance near zero bias is absent from the GS calculation. It may well be that GS simply did not investigate small enough bias currents to discover this resonance.

\begin{figure}[b]
\includegraphics{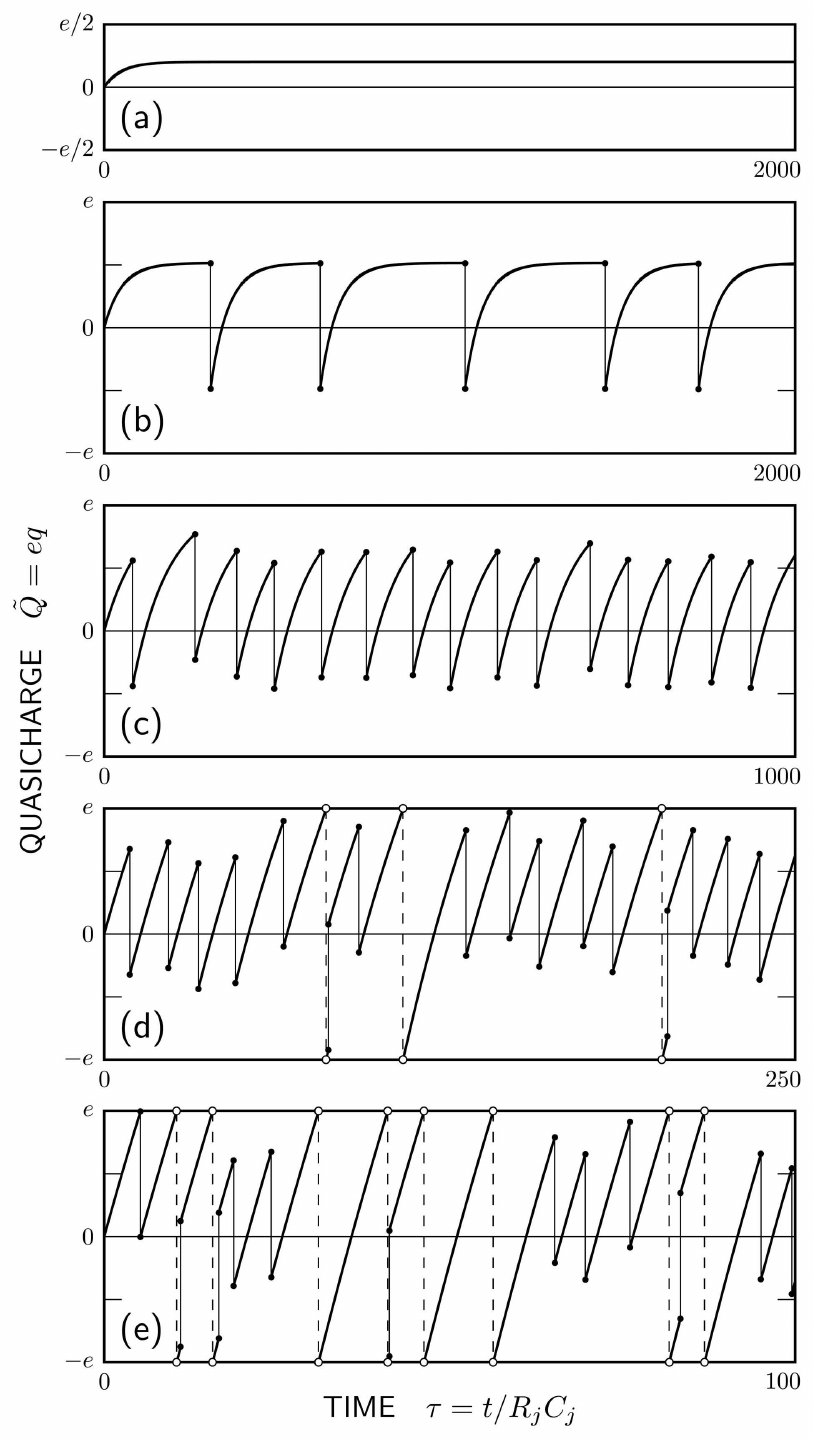}
\caption{\label{fig:qt1} Quasicharge as a function of time at five dc bias points chosen from the $\langle v\rangle$--$i_0$ curve of Fig.~\ref{fig:dciv1}: (a) $i_0=0.004$, (b) $i_0=0.0102$, (c) $i_0=0.02$, (d) $i_0=0.08$, and (e) $i_0=0.2$. In each instance, the system is initialized in the first band with zero quasicharge $(i_b,q)=(1,0)$. Single-electron tunneling events are shown by narrow vertical lines with a dot at each end, and Bloch reflections by dashed vertical lines with an open circle at each end.}
\end{figure}

To understand the nature of the motion represented in this $\langle v\rangle$--$i_0$ curve, it is useful to examine the detailed behavior of the quasicharge as a function of time. This is revealed in Fig.~\ref{fig:qt1}, where we plot $\tilde Q$ versus $\tau$ for the five bias points indicated by open circles in Fig.~\ref{fig:dciv1}. The key to understanding these plots is the simple form assumed by the single-electron tunneling rate at zero temperature.
\begin{equation}
\gamma=\left\{ \begin{array}{ll}
|\Delta\varepsilon|/2&\;\;\Delta\varepsilon\leq0\\
0&\;\;\Delta\varepsilon>0\end{array}\right.
\;\;\;\;\;\;\; (t_j=0)
\end{equation}
 In this case, tunneling is possible only when the energy $\varepsilon_1(\tilde Q)$ of the initial state is higher than that $\varepsilon_1(\tilde Q\pm e)$ of the final state, and inspection of Fig.~\ref{fig:energy}(a) reveals that this results only for initial states with quasicharge in the range $e/2<|\tilde Q|\leq e$. Conversely, single-electron tunneling is forbidden when $|\tilde Q|\leq e/2$.
 
 In these calculations, the quasicharge $q$ is updated during a given time step, say from $\tau$ to $\tau+\Delta\tau$, using a  fixed-step fourth-order Runge--Kutta algorithm to integrate Eq.~(\ref{eq:dq}). Simultaneously, we integrate $\gamma$ to determine the probability $P_e$ of single-electron tunneling during the interval, according to
 \begin{equation}
 P_e=1-\exp\left[\textstyle{-\int_{\tau}^{\tau+\Delta\tau}\gamma d\tau}\right].
 \end{equation}
We then select a random number $r$, uniformly distributed on the interval $(0,1)$, and if $r>P_e$, we assume that tunneling did not occur and proceed to the next integration step. But if $P_e\ge r$, we assume that tunneling occurred and add $\pm1$ to $q$ before continuing.

 Consider first the result for $i_0=0.004$, shown in Fig.~\ref{fig:qt1}(a). As with all of the curves in Fig.~\ref{fig:qt1}, the state of the system at $\tau=0$ is assumed to be $(i_b,\tilde Q)=(1,0)$. With time, the bias current begins to charge the junction capacitance $C_j$, initially raising its voltage $V$ rapidly then ever more slowly as the current $V/R_s$ is diverted through  the source resistance $R_s$. As a result, $V$ asymptotically approaches $\langle V\rangle=R_sI_0$ and (assuming $V\simeq\tilde Q/C_j$ for $|\tilde Q|\lesssim e/2$) the quasicharge approaches $\tilde Q=R_sC_jI_0$, both with a time constant $R_sC_j$.  In terms of dimensionless quantities,  the quasicharge $q$ approaches its approximate asymptote $i_0/g_s=0.2$ with a time constant of $1/g_s=50$. Because $q$ is always less than $0.5$, single-electron tunneling does not occur, and the nanoscale junction behaves as a simple capacitor. And from the asymptotic relation $\langle v\rangle=i_0/g_s$, we see that the initial slope of the $\langle v\rangle$--$i_0$ curve in Fig.~\ref{fig:dciv1} is $1/g_s$.

 When $i_0$ is increased to $0.01$ the asymptotic quasicharge reaches $q\simeq i_0/g_s=0.5$, and higher bias levels are sure to produce single-electron tunneling. Thus, at $i_0=0.0102$, after the quasicharge exceeds $q=0.5$ and $\varepsilon_1(q)>\varepsilon_1(q-1)$, single electron tunneling becomes possible, and the quasicharge is likely to jump suddenly from a value slightly greater than $0.5$ to a value slightly greater than $-0.5$, as shown in Fig.~\ref{fig:qt1}(b). When $C_j$ is discharged by such a tunneling event, the bias current immediately begins charging it again, and the process repeats at irregular intervals that reflect the random, Poisson character of the tunneling process. Despite the quantum nature of these oscillations, they are analogous to those of a classical relaxation oscillator. Finally, with the capacitor repeatedly discharged in this way, the average voltage drops from its peak of $\langle v\rangle\simeq0.5$ at $i_0\simeq0.01$ to $\langle v\rangle=0.355$ at $i_0=0.0102$.

With increasing bias above $i_0=0.0102$, the capacitor charges more rapidly and the asymptotic voltage and quasicharge generally increase. However, for $\varepsilon_j=0.2$ the voltage in the first band reaches a maximum of $v_{\rm max}=0.804$ at $q=0.874$ (see Fig.~\ref{fig:energy}(b)). As a result, for $i_0>g_sv_{\rm max}=0.0161$, there is no longer an asymptotic value of $v$ for which $dq/d\tau=i_0-g_sv=0$, and $q$ can in principle increase indefinitely.
 For the case $i_0=0.02$ shown in Fig.~\ref{fig:qt1}(c), however, the quasicharge simply oscillates between minimum values in the range $(-0.5,0.5)$ and maximum values in the range $(0.5,1)$.  These relaxation oscillations are relatively rapid because the charging rate is high and because the tunneling rate increases as $q$ approaches 1. (Note the change of time scale in the final frames of Fig.~\ref{fig:qt1}.) In fact, the tunneling rate is high enough that, while $q$ could increase beyond $1$, tunneling is overwhelmingly likely to occur first.

At yet higher dc bias and faster charging rates, however, $q$ exceeds $1$ on a regular basis. If we choose to restrict $q$ to the first Brillouin zone ($-1\le q\le1$), then when $q$ reaches $1$, we immediately reset $q$ to $-1$, an equivalent point in the energy band. In this case we say that the quasicharge has undergone a Bloch reflection and associate the event with coherent pair tunneling. In the extended zone scheme, on the other hand, $q$ is allowed to exceed $1$, and nothing of special significance occurs at $q=1$. In Fig.~\ref{fig:qt1}, we have chosen to restrict $q$ to the first Brillouin zone, and in frame (d) for $i_0=0.08$, we find Bloch reflections at three points. At these points, when $q$ reaches $1$, it is instantly reset to $-1$ before integration proceeds, and the jump is indicated by a dashed vertical line. In two of these cases, for $\tau$ near $80$ and $202$, the Bloch reflection is closely followed by single-electron tunneling from $q$ slightly greater than $-1$ to $q$ slightly greater than $0$. Single-electron tunneling is allowed here because $\varepsilon_1(q)>\varepsilon_1(q+1)$ for $q$ in the range $(-1,-0.5)$.

The average voltage increases between points (c) and (d) in Fig.~\ref{fig:dciv1} because the relaxation oscillations gradually shift to higher quasicharge as the charging rate increases with $i_0$. However, $\langle v\rangle$ reaches a peak at (d) due to the onset of Bloch reflections, which force the system to spend more time in negative charge states. Thus, the rounded peak in average voltage at (d) is usually referred to as the ``Bloch nose''. As the bias increases above that at (d), Bloch reflection becomes more frequent, and the average voltage falls as negative charge states are visited more often. This effect is apparent in Fig.~\ref{fig:qt1}(e) for $i_0=0.2$

\subsection{rf Bias}

\begin{figure}[b]
\includegraphics{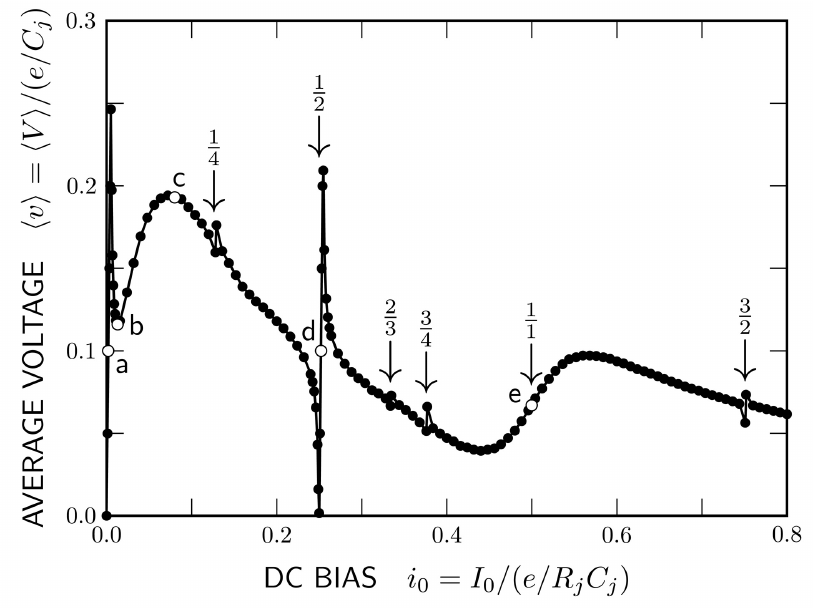}
\caption{\label{fig:rfiv1} Average voltage as a function of dc bias for the same set of parameters as Fig.~\ref{fig:dciv1} with an RF bias included ($\varepsilon_j=0.2$, $t_j=0$, $\alpha=0$, $g_s=0.02$, $i_1=0.4$, and $\omega=\pi/2$), computed by Monte Carlo simulation with an averaging time of $10^6$ rf drive cycles. Arrows labeled $\frac{n}{m}$ mark current steps at which $m$ Bloch oscillations are nominally completed during $n$ drive cycles.}
\end{figure}

When an rf bias is included in our example system, we obtain the voltage--current curve shown in Fig.~\ref{fig:rfiv1} for $i_1=0.4$ and $\omega=\pi/2$. This curve is in good agreement with that for the same parameter set shown in Fig.~8 of GS, although GS does not include points at small enough dc bias to reveal the spike near $i_0=0$. In particular, we find the same sharp step at $i_0=0.25$, corresponding to $I_0=ef$, and the same broad step at $i_0=0.5$, corresponding to $I_0=2ef$, that were observed in GS. These steps reveal the tendency of Bloch oscillations to phase lock with the applied rf bias and are the basis for the proposed current standard.

The step at $i_0=0.5$ represents synchronized motion in which one Bloch oscillation occurs during each rf cycle, while for that at $i_0=0.25$ a Bloch oscillation is completed only after two rf cycles. The former is an example of harmonic phase lock in which $n$ Bloch oscillations are completed during $m=1$ drive cycles, while the latter is a case of subharmonic phase lock, in which $n$ oscillations are completed during $m\geq2$ drive cycles. In either case, the step nominally occurs at $I_0=(n/m)2ef$ and defines an approximate quantized current. However, we are naturally led to ask why the harmonic step with $n/m=1/1$ is so much wider than the subharmonic steps with $n/m=1/4$, $1/2$, $2/3$, $3/4$, and $3/2$.

\begin{figure}[b]
\includegraphics{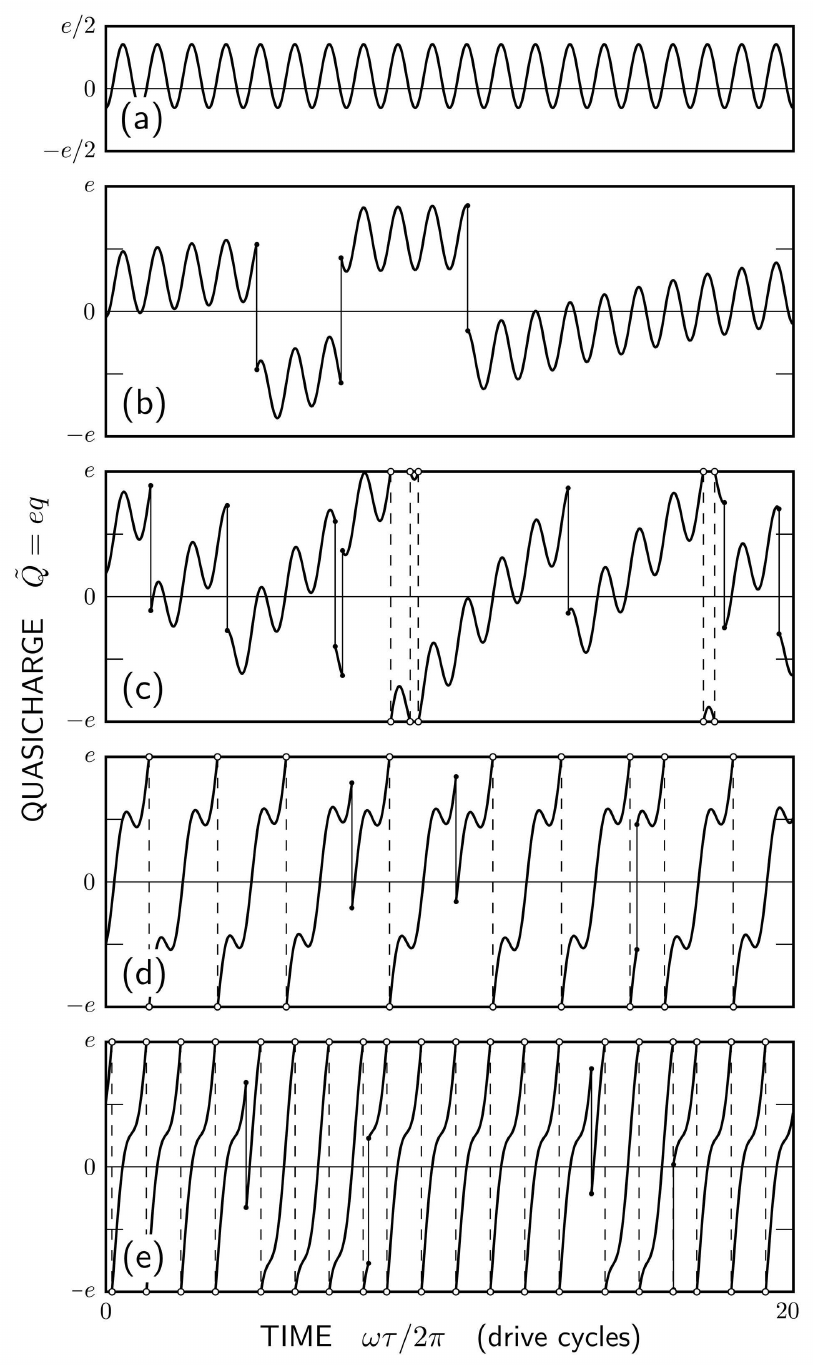}
\caption{\label{fig:qt2} Quasicharge as a function of time at five dc bias points chosen from the $\langle v\rangle$--$i_0$ curve of Fig.~\ref{fig:rfiv1}: (a) $i_0=0.002$, (b) $i_0=0.013$, (c) $i_0=0.08$, (d) $i_0=0.252$, and (e) $i_0=0.5$. In each instance, the equation of motion is integrated for at least 200 drive cycles to eliminate transients before plotting 20 cycles of representative motion. Single-electron tunneling events are shown by narrow vertical lines with a dot at each end, and Bloch reflections by dashed vertical lines with an open circle at each end.}
\end{figure}

Before attempting to answer this question, we explore the nature of the motion at each of the five bias points marked by open circles in Fig.~\ref{fig:rfiv1}. Consider first the case shown in Fig.~\ref{fig:qt2}(a) for $i_0=0.002$. Here $q$ is always less than $0.5$, so $q\simeq v$, and in the sinusoidal steady state we expect
\begin{equation}
q\simeq v=i_0/g_s+i_1\sin(\omega\tau+\phi)/\sqrt{g_s^2+\omega^2}\label{eq:ss}.
\end{equation}
Thus, as shown, $\tilde Q$ oscillates with an amplitude of $0.255~e$ about an average value of $0.1~e$, uninterrupted by single-electron tunneling.

If Eq.~(\ref{eq:ss}) is applied to the case of $i_0=0.013$  in Fig.~\ref{fig:qt2}(b), it implies a steady state with an average quasicharge of $\langle\tilde Q\rangle=0.65~e$, which exceeds the threshold for single-electron tunneling, and a peak voltage of $v=0.905$, which exceeds the threshold of  $v_{\rm max}=0.804$ for Bloch reflection. Thus $\tilde Q$ fails to assume a steady state in Fig.~\ref{fig:qt2}(b) and instead displays roughly sinusoidal behavior interrupted at intervals by single-electron tunneling and very occasional Bloch reflections (the latter do not appear in Fig.~\ref{fig:qt2}(b)). At yet higher dc bias, Bloch reflections become more frequent ({\it cf.} Fig.~\ref{fig:qt2}(c)) and lead to the Bloch nose at bias point (c) in Fig.~\ref{fig:rfiv1}.

The final bias points, (d) and (e) in Fig.~\ref{fig:rfiv1}, are centered on the steps at $I_0=ef$ and $I_0=2ef$.  As expected for subharmonic and harmonic phase locking with $n/m=1/2$ and $n/m=1/1$, the corresponding quasicharge curves in Fig.~\ref{fig:qt2} reveal significant intervals during which Bloch reflections occur periodically, with a period of two drive cycles in (d) and one drive cycle in (e). In both cases, however, these patterns are interrupted at irregular intervals by single-electron tunneling. When tunneling occurs for initial quasicharge in the range $e/2<\tilde Q<e$ the time between Bloch reflections is lengthened, while for initial quasicharge in the range $-e<\tilde Q<-e/2$ the time between reflections is shortened. Because the relative proportions of these competing events and the resulting average voltages change gradually with dc bias, there is no signature in the $\langle v\rangle$--$i_0$ curve that identifies the exact dc bias at which $I_0=ef$ or $2ef$.

\begin{figure}[t]
\includegraphics{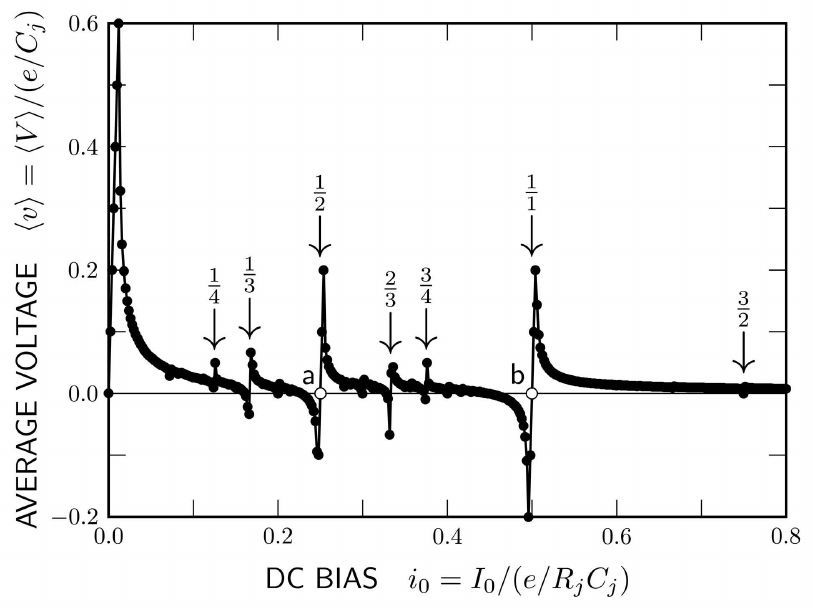}
\caption{\label{fig:rfivne} Average voltage as a function of dc bias in the absence of single-electron and Zener tunneling ($\varepsilon_j=0.2$, $t_j=0$, $\alpha=0$, $G_s=0.02$, $i_1=0.4$, and $\omega=\pi/2$), computed by Monte Carlo simulation with an averaging time of $10^5$ rf drive cycles. Arrows labeled $\frac{n}{m}$ mark current steps at which $m$ Bloch oscillations are nominally completed during $n$ drive cycles.}
\end{figure}

\begin{figure}[b]
\includegraphics{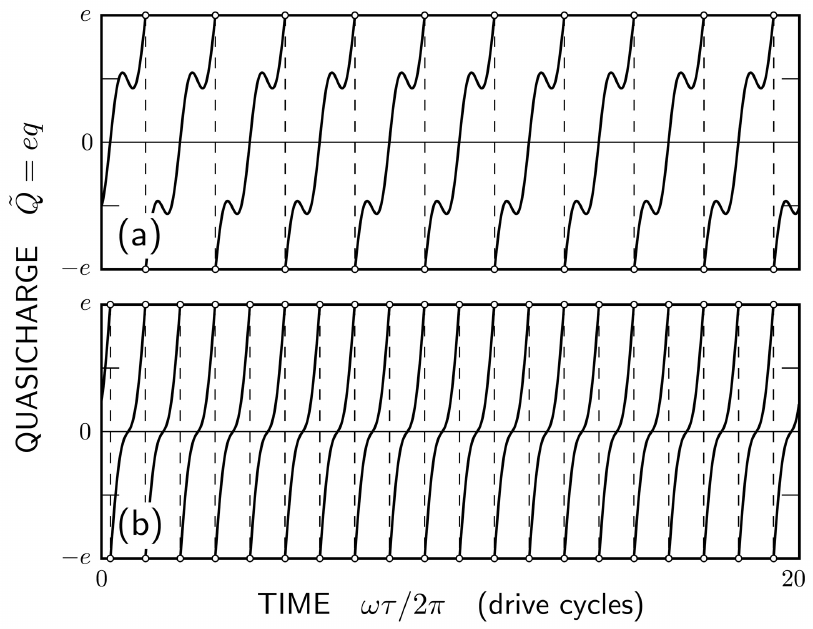}
\caption{\label{fig:qt3} Quasicharge as a function of time at two dc bias points chosen from the $\langle v\rangle$--$i_0$ curve of Fig.~\ref{fig:rfivne}: (a) $i_0=0.25$ and (b) $i_0=0.5$. In each instance, the equation of motion is integrated for at least $200$ drive cycles to eliminate transients before plotting $20$ cycles of steady-state motion. Bloch reflections are represented by dashed vertical lines with an open circle at each end.}
\end{figure}

 A qualitative difference between the steps at (d) and (e) in Fig.~\ref{fig:rfiv1} is evident in the slope of the $\langle v\rangle$--$i_0$ curve in the two cases, with $d\langle v\rangle/di_0=50$ or $d\langle V\rangle/dI_0=R_s$ at (d) and $d\langle v\rangle/di_0\simeq0.8$ or $d\langle V\rangle/dI_0\simeq 0.8R_j$ at (e). This approximate correspondence between the slopes of the two steps and the resistances $R_s$ and $R_j$ suggests that the feedback mechanism that creates phase lock is distinctly different for subharmonic and harmonic steps. Further insight into this possibility results from examining the $\langle v\rangle$--$i_0$ curve for the same parameters as Fig.~\ref{fig:rfiv1} but with single-electron tunneling completely suppressed. Results for this case are shown in Figs.~\ref{fig:rfivne} and \ref{fig:qt3}.

As seen in Fig.~\ref{fig:rfivne}, in the absence of single-electron tunneling the Bloch nose is eliminated from the voltage--current curve, and harmonic phase lock at $i_0=0.5$ gives rise to the same sharp step as previously observed for subharmonic steps. The steady-state motion for $n/m=1/2$ and $1/1$ is shown in frames (a) and (b) of Fig.~\ref{fig:qt3}, where we see phase lock uninterrupted by tunneling. Comparing these plots with frames (d) and (e) of Fig.~\ref{fig:qt2} reveals what might be a critical difference between subharmonic and harmonic phase locking in the presence of tunneling. For $n/m=1/2$, we see in Fig.~\ref{fig:qt2}(d) that tunneling shifts $q$ by $\pm1$, but because the drive repeats itself twice between Bloch reflections, this shift allows the oscillation pattern to immediately resume the steady-state motion of Fig.~\ref{fig:qt3}(a). In this case, Bloch reflection is advanced or delayed by tunneling, but tunneling doesn't upset the pattern of phase lock. On this subharmonic step, regardless of the presence or absence of tunneling, lock results from feedback through $R_s$, and the system relaxes to locked motion with a characteristic time of $R_sC_j=12.5$ drive cycles.

In contrast, for the harmonic step at $i_0=0.5$ we find that single-electron tunneling completely upsets the pattern of phase lock found in the absence of tunneling. This disruption becomes evident if we focus on an inflection point in the $q$ versus $\tau$ curve that in the absence of tunneling occurs at $q=0$, as seen in Fig.~\ref{fig:qt3}(b). This is the expected location of the inflection point when phase lock is established by feedback through the source resistance $R_s$. Examining $q$ versus $\tau$ with tunneling present, as shown in Fig.~\ref{fig:qt2}(e), however, we find that the inflection point alternates irregularly between $q\simeq 0.24$ and $q\simeq-0.76$, as the junction is buffeted by tunneling events. (Over longer periods of time, the inflection point can be found at any value of quasicharge in the range $-1<q<1$.) This erratic behavior results because there is only one drive cycle between Bloch reflections on the $n/m=1/1$ step, so when tunneling shifts $q$ by $\pm1$, the quasi charge jumps to a value that would otherwise occur a half drive cycle later. Thus, rather than jumping a full drive cycle as on the $1/2$ step,  a shift in $q$ by $\pm1$ has the effect of jumping a half drive cycle on the $1/1$ step. Such a half-cycle jump puts the system far from the phase-lock state associated with feedback through $R_s$, and because the relaxation time required to regain lock by this mechanism ($12.5$ drive cycles)  is far longer than the time between tunneling events, feedback through $R_s$ is not effective in maintaining lock on the $1/1$ step.

These arguments lead to the conclusion that phase lock on the harmonic step probably results from feedback through the junction resistance $R_j$, in spite of the fact that the current through this element is the shot noise of single-electron tunneling. That is, much like electronic systems that use impulsive feedback as a means of control, phase lock on the $1/1$ step is apparently maintained by single-electron tunneling. This conclusion is confirmed by three observations. First, the rate of single-electron tunneling is a strong function of $q$, so tunneling is not entirely random and is capable of providing feedback. Second, the slope of the $1/1$ step is of order $R_j$,\cite{gei88} as expected if $R_j$ provides the feedback to create the step. Third, when the $\langle v\rangle$--$i_0$ curve is computed without a source resistance ($g_s=0$), the narrow subharmonic steps in Fig.~\ref{fig:rfiv1} are eliminated, while the Bloch nose and the broad harmonic step at $I_0=2ef$ are retained. As will be seen later, the higher-order harmonic $n/1$ steps at $I_0=2nef$ also have slopes of order $R_j$ and are also created by the feedback from single-electron tunneling. Thus, in any system with non-zero single-electron tunneling, the principal Bloch steps at $I_0=2nef$, the epitome of coherent pair tunneling, ironically owe their existence to single-electron tunneling, the very process that introduces errors into the quantization implied by $I_0=2nef$.

\section{Ensemble Simulation}

 In a Monte Carlo calculation, the state $(i_b,q)$ of a single junction is tracked over a long period of time, and $\langle v\rangle$ is evaluated as a time average. However, we can also consider an ensemble of identical systems with random initial conditions and calculate the steady-state probability density $\rho_{i_b}(q)$ derived from a Langevin equation. In this case $\langle v\rangle$ is computed as an ensemble average over $i_b$ and $q$ and a time average over one drive cycle. The ensemble approach has the advantage of computational speed, because averaging over long time periods is not required, but it is less efficient in cases where the probability distribution is limited to a narrow range of quasicharge.

\begin{figure}[t]
\includegraphics{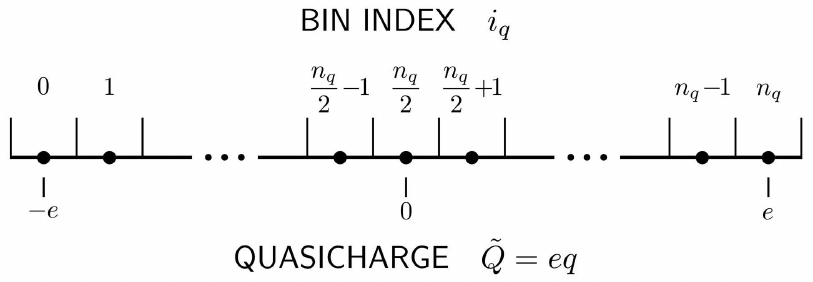}
\caption{\label{fig:bin} Quasicharge bin index as a function of quasicharge for a typical band. The number $n_q$ of bins per band is taken to be even, so the bin with index $i_q=n_q/2$ is centered at $\tilde Q=0$. Bin $0$ at $\tilde Q=-e$ is equivalent to bin $n_q$ at $\tilde Q=e$ and is omitted from the count.}
\end{figure}

 Rather than begin with an equation of motion for $\rho_{i_b}(q)$, we will instead simply describe how our computer program works based on Eqs.~(\ref{eq:v})--(\ref{eq:PZ1}). This will afford a more definite and perhaps clearer picture of exactly how the system is modeled. To begin, we note that while $\rho_{i_b}(q)$ is nominally a continuous function of $q$ and $\rho_{i_b}(q)dq$ is the probability of finding the system in band $i_b$ with a quasicharge between $q$ and $q+dq$, a practical program results when we break the quasicharge into a finite number of bins and consider only the probabilities of finding the system in the various bins. The indexing scheme for the quasicharge bins of one band is shown in Fig.~\ref{fig:bin}. Here, a bin with index $i_q$ is associated with the quasicharge at the center of the bin according to $\tilde Q(i_q)= eq(i_q)=e(2i_q/n_q-1)$, where $n_q$ is the number of bins allocated to each band, and we allow a maximum of $n_b$ bands. Our program focuses on computing the probability $P_s(i_b,i_q)$ of finding the system in bin $i_q$ of band $i_b$, and these bin and band indices completely define the state of the system within the resolution of the calculation.

The dynamics of this discrete-quasicharge approximation to the system is most easily expressed by introducing an index $i_s$,
\begin{equation}
i_s=n_q(i_b-1)+i_q,\;\;\;\;\;(i_b\leq n_b,\; i_q\leq n_q)
\end{equation}
which combines $i_b$ and $i_q$ to specify one of the $n_bn_q$ states. The probabilities $P_s(i_s)$ for occupying the states $i_s$ evolve in time according to the master equation,
\begin{equation}
\frac{dP_s(i_s)}{d\tau}=\sum_{i_s'=1}^{n_bn_q}A(i_s,i_s')P_s(i_s'),\label{eq:dP}
\end{equation}
where the matrix $A(i_s,i_s')$ specifies the rate at which probability in state $i_s'$ is transferred to state $i_s$ per unit probability in state $i_s'$,  as determined by Eqs.~(\ref{eq:dq})--(\ref{eq:PZ1}).

While by far the majority of the $(n_bn_q)^2$ elements of the rate matrix $A$ are zero, precisely specifying all of the nonzero elements is an exercise in conditional statements best left to a computer program. However, we will examine the general nature of the terms contributed by the three processes specified by Eqs.~(\ref{eq:dq})--(\ref{eq:PZ1}). Before doing so, it is useful to introduce the notation,
\begin{equation}
A(i_b,i_q; i_b',i_q')=A(n_q(i_b-1)+i_q, n_q(i_b'-1)+i_q'),
\end{equation}
which allows elements of the rate matrix to be identified by the physically relevant band and quasicharge indices.

Equation~(\ref{eq:dq}) tells us that a junction current $i_j>0$ has the effect of shifting probability from bin $i_q'$ to the adjacent bin $i_q=i_q'+1$ within the same band. More specifically, since each bin has a width $\Delta q=2/n_q$, the time $\Delta\tau$ required to shift all of the probability in one bin to an adjacent bin is $\Delta\tau=\Delta q/|i_j|=2/n_j|i_j|$, and the rate per unity probability is $1/\Delta\tau=n_j|i_j|/2$. That is,
\begin{equation}
A(i_b',i_q'+1;i_b',i_q')=\frac{n_j}{2}|i_j|,\label{eq:A1}
\end{equation}
where,
\begin{equation}
i_j =i_0+i_1\sin(\omega\tau)-g_sv_{i_b'}[q(i_q')].
\end{equation}
Here it is understood that $\pm n_q$ is added to the final quasicharge index $i_q$ as required to keep it within the range $1\le i_q\le n_q$. Thus for an initial quasicharge $i_q'=n_q$ in Eq.~(\ref{eq:A1}), the final quasicharge is $i_q=1$ rather than $n_q+1$. An equation similar to Eq.~(\ref{eq:A1}) results for $i_j<0$, except that probability is shifted to an adjacent bin of lower rather than higher quasicharge: $i_q=i_q'-1$.

The rate of single-electron tunneling given by Eq.~(\ref{eq:gamma}) translates directly into elements of the rate matrix that generally connect a bin $i_q'$ in a given band $i_b'$ to a bin $i_q=i_q'\pm n_q/2$ in another band either just above or just below the given band, $i_b=i_b'\pm1$, although tunneling can also occur within band $1$. The shift in $i_q'$ by $\pm n_q/2$ assures that the quasicharge $\tilde Q$ changes by $\pm e$. Single-electron tunneling to the next higher band is governed by a matrix element of the form,
\begin{equation}
A(i_b'+1,i_q'\pm n_q/2;i_b',i_q')=\frac{\Delta\varepsilon/2}{\exp(\Delta\varepsilon/t_j)-1},
\end{equation}
where the difference in energy $\Delta\varepsilon$ between the final and initial states is
\begin{equation}
\Delta\varepsilon= \varepsilon_{i_b'+1}\left[q(i_q'\pm n_q/2)\right]-
\varepsilon_{i_b'}\left[q(i_q')\right],
\end{equation}
and similar formulas result for other possibilities.

Finally, we need to account for Zener tunneling between bands, which can occur when the quasicharge passes through an energy maximum or minimum that brings it close to a second band, either at $\tilde Q=0$ or $e$. Consider, for example, the possibility of Zener tunneling from band $1$ to band $2$ at $\tilde Q=e$ with $d\tilde Q/dt=I_j>0$. According to Eq.~(\ref{eq:PZ1}) this will occur with probability
\begin{equation}
P_{Z,1\leftrightarrow2}=\exp\left[-\frac{(\Delta\varepsilon)^2}{4\alpha|i_j|}\right],
\end{equation}
where $\Delta\varepsilon$ is the energy gap between the first and second bands,
\begin{equation}
\Delta\varepsilon=\varepsilon_2[q(n_q)]-\varepsilon_1[q(n_q)].
\end{equation}
To incorporate this tunneling event into the rate matrix, we assume that it occurs as probability is shifted by the drive current $i_j>0$ from bin $i_q'=n_q-1$ of band $i_b'=1$, with the probability ending up either in bin $i_q=n_q$ of band $i_b=2$ with probability $P_{Z,1\leftrightarrow2}$ or in bin $i_q=n_q$ of band $i_b=1$ with probability $1-P_{Z,1\leftrightarrow2}$. Thus, Zener tunneling can be included by replacing the matrix element $A(1,n_j;1,n_j-1)$ given by Eq.~(\ref{eq:A1}) with the pair of matrix elements,
\begin{eqnarray}
A(1,n_q;1,n_q-1)&=&(1-P_{Z,1\leftrightarrow2})\frac{n_q|i_j|}{2},\\
A(2,n_q;1,n_q-1)&=&P_{Z,1\leftrightarrow2}\frac{n_q|i_j|}{2},
\end{eqnarray}
where the current $i_j$ is
\begin{equation}
i_j=i_0+i_1\sin(\omega\tau)-g_sv_1[q(n_q-1)].
\end{equation}
Matrix elements for other Zener tunneling events occurring at $i_q=n_q/2$, from higher to lower bands, or with negative $i_j$, can be constructed in a similar fashion.

All of the matrix elements $A(i_s,i_s')$ discussed above define the positive rate at which probability flows from state $i_s'$ to another state $i_s$. However, in order to conserve probability, we must deduct this probability flow from the state of origin $i_s'$. Thus, the diagonal elements of the rate matrix are given by,
\begin{equation}
A(i_s',i_s')=-\sum_{\scriptstyle i_s=1\atop\scriptstyle i_s\neq i_s'}^{n_bn_q}A(i_s,i_s'),
\label{eq:con}
\end{equation}
and this formula completes our explication of the rate matrix.

\subsection{dc Bias}

As an example of the ensemble approach to calculating $\langle v\rangle$--$i_0$ curves, we turn again to the case considered in Fig.~\ref{fig:dciv1} for dc bias only. Without an rf bias, the system is expected to approach a steady state in which the bin probabilities are independent of time and $dP_s(i_s)/d\tau=0$ for all quasicharge bins. According to Eq.~(\ref{eq:dP}), this steady state is defined by
\begin{equation}
\sum_{i_s'=1}^{n_bn_q}A(i_s,i_s')P_s(i_s')=0,\label{eq:Ass}
\end{equation}
which provides a system of $n_bn_q$ linear equations for the $n_bn_q$ bin probabilities $P_s(i_s')$. However, conservation of probability, Eq.~(\ref{eq:con}), implies that any one of these equations is a linear combination of the other $n_bn_q-1$ equations. To obtain a full set of $n_bn_q$ independent equations, we replace one equation of the above set with the normalization condition,
\begin{equation}
\sum_{i_s'=1}^{n_bn_q}P_s(i_s')=1\label{eq:norm}.
\end{equation}
When combined, Eqs.~(\ref{eq:Ass}) and (\ref{eq:norm}) allow a direct calculation of the bin probabilities in the case of dc bias only. The average voltage can then be evaluated according to,
\begin{equation}
\langle v\rangle=\sum_{i_s=1}^{n_bn_q} P_s(i_s)v(i_s),\label{eq:vav}
\end{equation}
where it is understood that $v(i_s)$ is the voltage $v_{i_b}(i_q)$ of the band $i_b$ and bin $i_q$ corresponding to the state $i_s$.

\begin{figure}[b]
\includegraphics{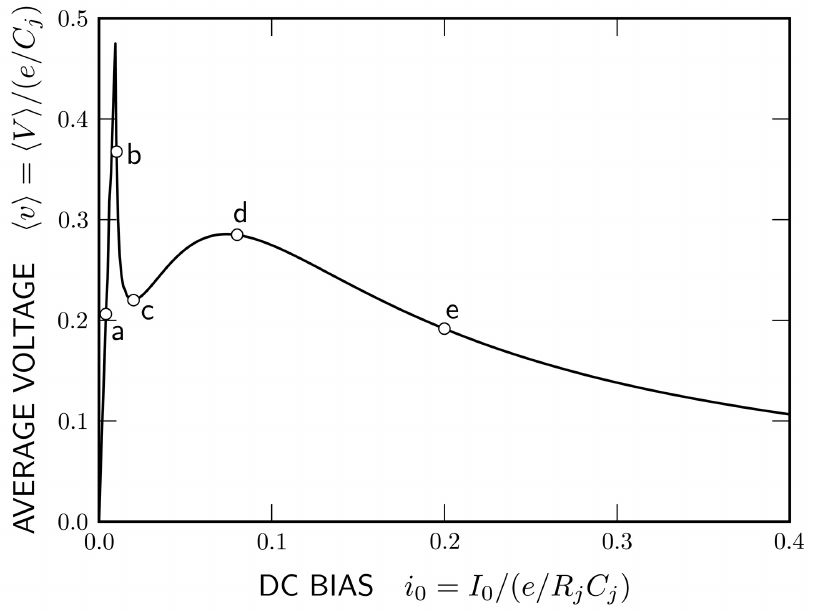}
\caption{\label{fig:dciv2} Average voltage as a function of dc bias for $\varepsilon_j=0.2$, $t_j=0$, $\alpha=0$, $g_s=0.02$, and $i_1=0$ (the same set of parameters as the Monte Carlo calculations in Fig.~\ref{fig:dciv1}), computed as an ensemble average over $n_q=1000$ quasicharge bins. For $i_0\leq0.015$, bin probabilities were evaluated by allowing the system to relax to a steady state using Eq.~(\ref{eq:dP}), while for $i_0>0.015$ probabilities were obtained directly by solving a system of linear equations, Eqs.~(\ref{eq:Ass}) and (\ref{eq:norm}).}
\end{figure}

As shown in Fig.~\ref{fig:dciv2}, the $\langle v\rangle$--$i_0$ curve from our ensemble calculation closely matches that of Fig.~\ref{fig:dciv1}, calculated by a Monte Carlo methods. Actually, Fig.~\ref{fig:dciv2} includes data from two types of  calculation. For $i_0>0.015$ a calculation based on Eqs.~(\ref{eq:Ass}) and (\ref{eq:norm}) is efficient and accurate, but for $i_0<0.015$, where we find a spike in $\langle v\rangle$, this direct method often produces spurious results. Thus, for low bias, we have instead returned to Eq.~(\ref{eq:dP}) and, beginning with an initially uniform probability distribution, simply allowed the system to evolve in time until the distribution reaches a steady state. While this relaxation approach is less efficient than solving Eqs.~(\ref{eq:Ass}) and (\ref{eq:norm}), it converges relatively quickly (usually within a few $R_jC_j$ times) to an accurate distribution.

\begin{figure}[t]
\includegraphics{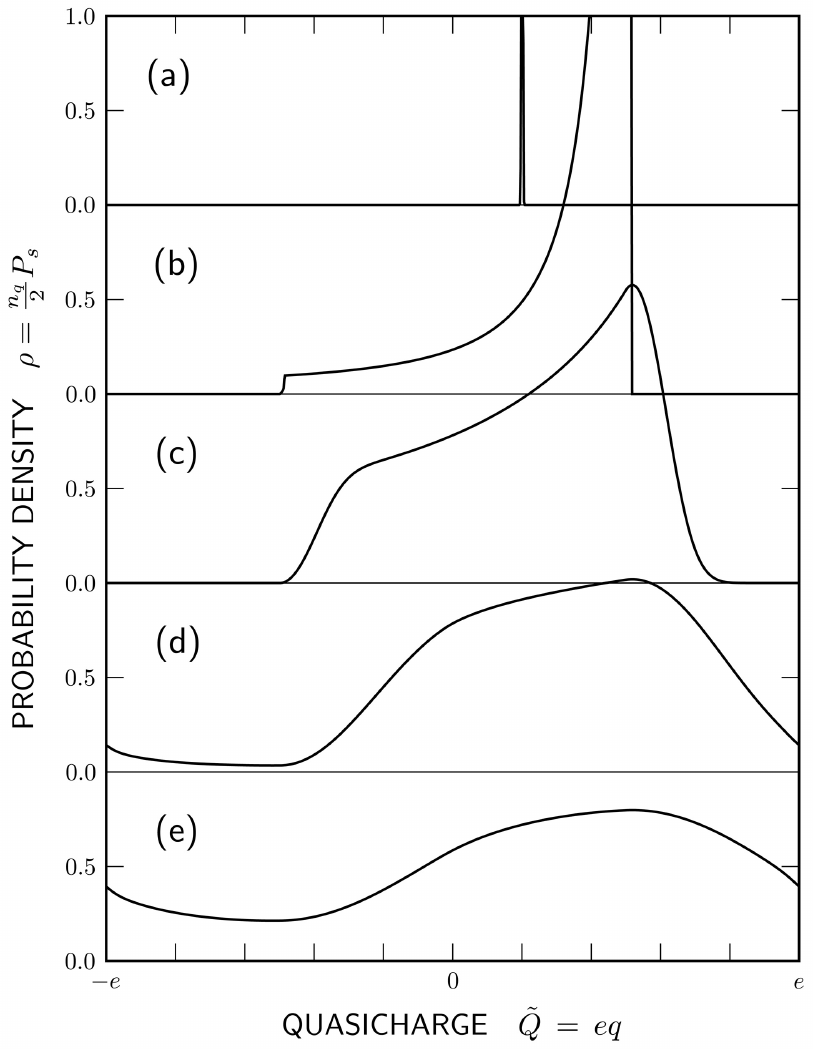}
\caption{\label{fig:dcrho} Steady-state probability density as a function of quasicharge at five dc bias points selected from the $\langle v\rangle$--$i_0$ curve of Fig.~\ref{fig:dciv2}: (a) $i_0=0.004$, (b) $i_0=0.0102$, (c) $i_0=0.02$, (d) $i_0=0.08$, and (e) $i_0=0.2$. These distributions can be compared directly with the corresponding Monte Carlo quasicharge versus time plots of Fig.~\ref{fig:qt1}.}
\end{figure}

To better understand the nature of ensemble calculations, we examine the probability density $\rho$ as a function of quasicharge, plotted in Fig.~\ref{fig:dcrho} for five bias points selected from Fig.~\ref{fig:dciv2}. These are the same bias points for which quasicharge is plotted as a function of time in Fig.~\ref{fig:qt1}, and it's not difficult to predict $\rho(\tilde Q)$ from $\tilde Q(\tau)$. At bias point (a), for example, the steady-state quasicharge is fixed at $\tilde Q\simeq 0.2~e$, so we expect the corresponding $\rho(\tilde Q)$ to include a delta-function at $0.2~e$. This expectation is fulfilled in Fig.~\ref{fig:dcrho}(a), where $\rho$ is off scale at $0.2~e$, and the raw data reveal two adjacent bins near $0.2~e$ that include $99~\%$ of the probability. Similarly, the quasicharge waveform in Fig.~\ref{fig:qt1}(b) shows that the system lingers near $\tilde Q=0.5~e$ but occasionally dips to roughly $-0.5~e$,  so we're not surprised to find a peak in $\rho$ near $\tilde Q=0.5~e$ and a probability tail that extends down to about $-0.5~e$, as shown in Fig.~\ref{fig:dcrho}(b). At yet higher dc biases, Bloch reflection becomes possible and the probability density is spread over the full range of quasicharge, from $-e$ to $e$, as in Figs.~\ref{fig:dcrho}(d) and (e).

In these ensemble calculations, the number of quasicharge bins was chosen to be $n_q=1000$, which allows sufficient resolution in $\tilde Q$ that the delta function in probability near $\tilde Q=0.2~e$ in Fig.~\ref{fig:dcrho}(a) is well resolved. However, in the absence of such sharp structure, as in Figs.~\ref{fig:dcrho}(d) and (e), fewer bins are required, and the probability distribution is usually represented accurately using just $100$ bins per band. In later simulations, we typically use this smaller number of bins.

\subsection{rf Bias}

While it may seem unlikely, the ensemble approach is also useful in the presence of an rf  bias.
In this case, the rate matrix $A$ is time dependent, and Eq.~(\ref{eq:Ass}) is no longer applicable, but the relaxation approach remains viable. This approach depends on the assumption that the probability density will relax to a steady-state function with the same periodicity as the rf drive. This assumption is confirmed by numerical simulations in which the $n_bn_q$ coupled probabilities $P_s(i_s)$ are calculated from Eq.~(\ref{eq:dP}) using a fourth-order Runge-Kutta algorithm. Because the probabilities typically converge to a periodic solution within a few rf drive cycles, the relaxation approach offers a practical method of computing $\langle v\rangle$--$i_0$ curves. In this case, $\langle v\rangle$ is an average over the ensemble and over time,
\begin{equation}
\langle v\rangle=
\sum_{i_s=1}^{n_bn_q}\frac{1}{\tau_p}\int_0^{\tau_p}P_s(i_s)v(i_s)d\tau,
\end{equation}
where $\tau_p=2\pi/\omega$ is the period of the rf drive. One precaution that must be taken in integrating Eq.~(\ref{eq:dP}) is choosing a time step $\Delta\tau$ small enough that probability is never driven by $i_j$ beyond the adjacent bin --- that is, $\Delta\tau<2/n_q|i_j|$.

\begin{figure}[t]
\includegraphics{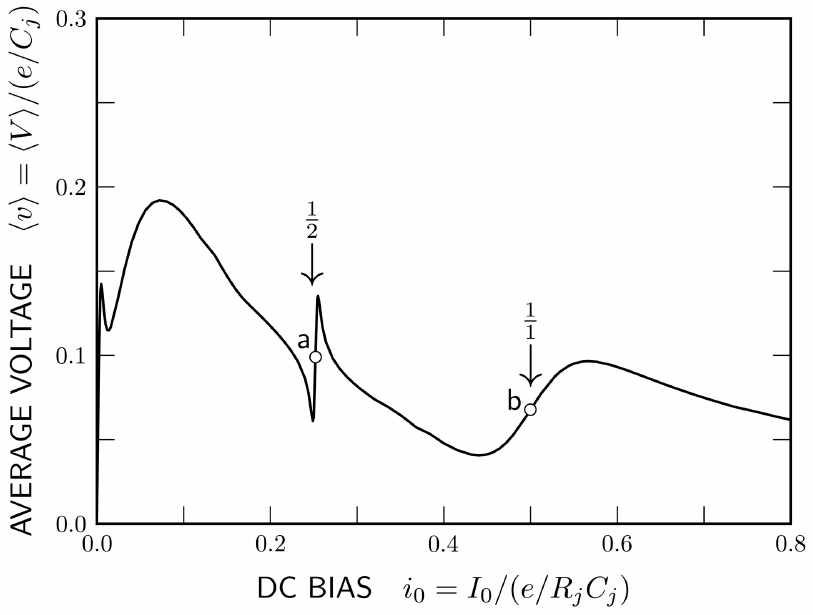}
\caption{\label{fig:rfiv2} Average voltage as a function of dc bias for $\varepsilon_j=0.2$, $t_j=0$, $\alpha=0$, $g_s=0.02$, $i_1=0.4$, and $\omega=\pi/2$ (the same set of parameters as the Monte Carlo calculations in Fig.~\ref{fig:rfiv1}), computed as an ensemble average over $n_q=1000$ quasicharge bins.}
\end{figure}

 A voltage--current curve for an rf-biased junction computed by the ensemble approach is shown in Fig.~\ref{fig:rfiv2} for the same case as evaluated by Monte Carlo simulation in Fig.~\ref{fig:rfiv1}. The striking difference between these curves is the degree to which the prominent subharmonic steps in Fig.~\ref{fig:rfiv1} are suppressed in Fig.~\ref{fig:rfiv2}. This loss of fine structure in the $\langle v\rangle$--$i_0$ curve is typical of ensemble calculations and probably derives from replacing the continuous quasicharge variable with discrete quasicharge bins. On the other hand, there is excellent agreement between the two calculations with regard to the Bloch nose at $i_0\simeq 0.08$ and the first harmonic step at $i_0=0.5$. Thus, ensemble calculations offer an efficient alternative to Monte Carlo simulations if fine structure, such as  subharmonic steps, is not of special interest.

\begin{figure}[b]
\includegraphics{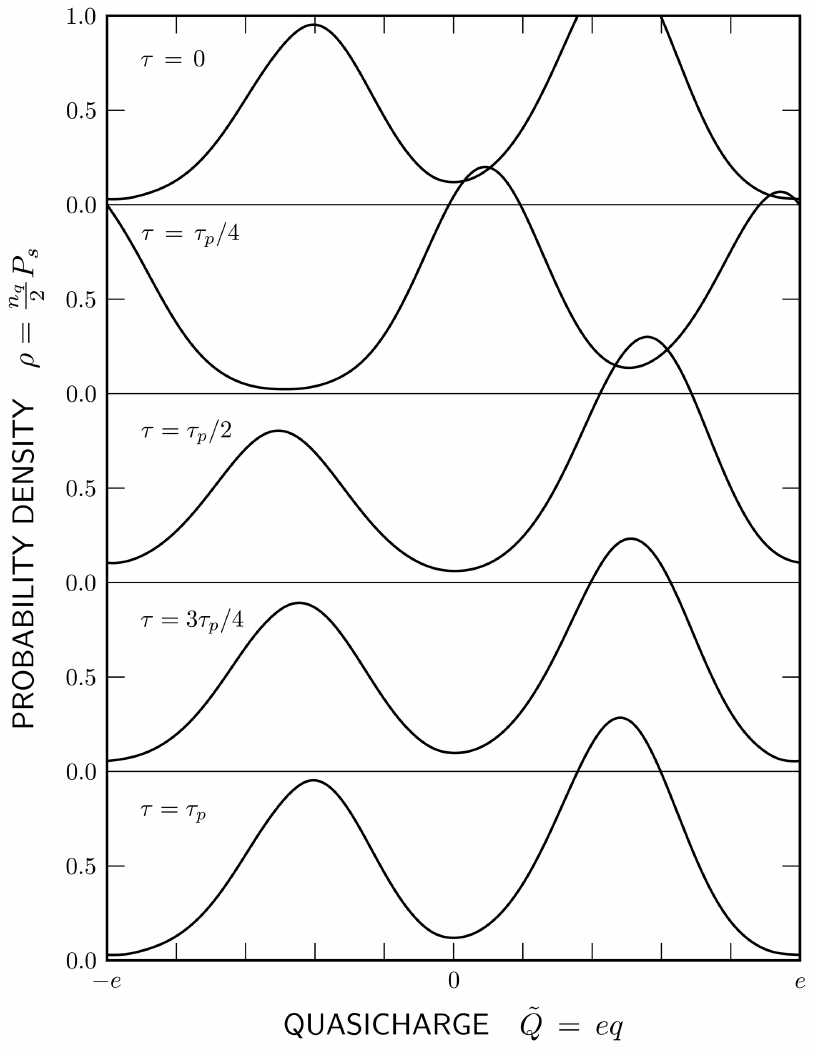}
\caption{\label{fig:rfrho1} Steady-state probability density as a function of quasicharge at five times during the rf drive cycle for bias point (a), $i_0=0.252$, of the $\langle v\rangle$--$i_0$ curve in Fig.~\ref{fig:rfiv2}. These distributions can be compared with the corresponding Monte Carlo quasicharge versus time plot of Fig.~\ref{fig:qt2}(d).}
\end{figure}

\begin{figure}[t]
\includegraphics{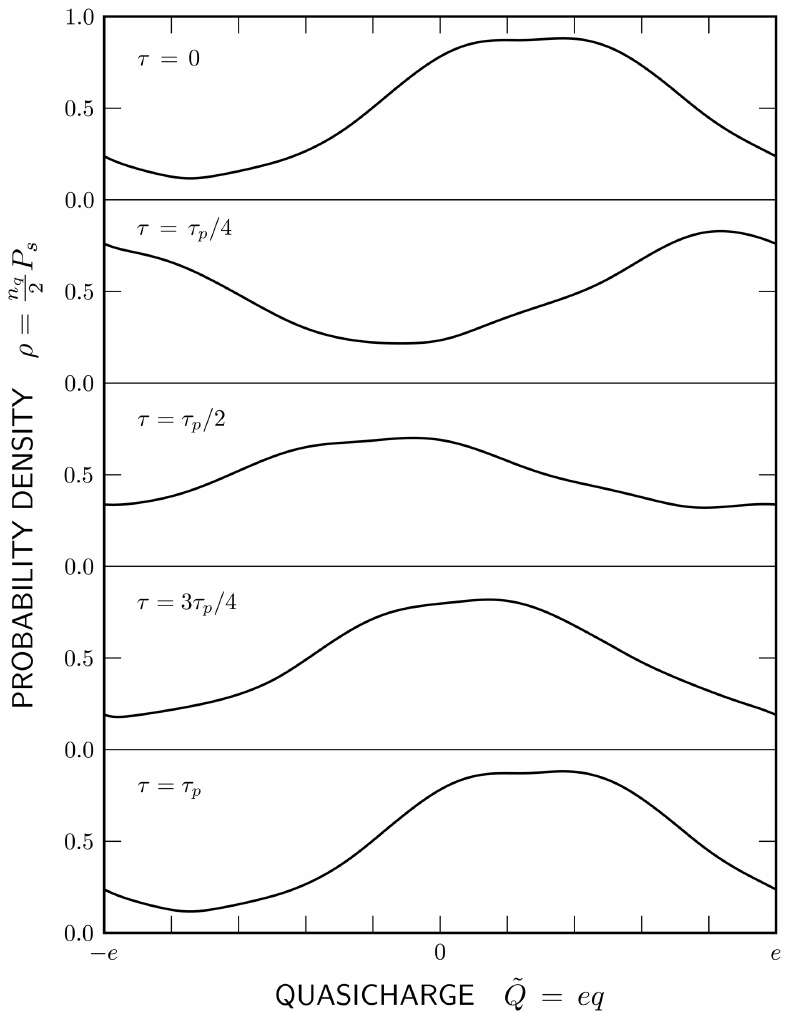}
\caption{\label{fig:rfrho2} Steady-state probability density as a function of quasicharge at five times during the rf drive cycle for bias point (b), $i_0=0.5$, of the $\langle v\rangle$--$i_0$ curve in Fig.~\ref{fig:rfiv2}. These distributions can be compared with the corresponding Monte Carlo quasicharge versus time plot of Fig.~\ref{fig:qt2}(e).}
\end{figure}

An idea of the inner workings of an ensemble calculation in the presence of an rf drive is given by plots of the probability density, shown in Figs.~\ref{fig:rfrho1} and \ref{fig:rfrho2} for the bias points (a) and (b) identified in Fig.~\ref{fig:rfiv2}. Consider first bias point (a) centered on the $n/m=1/2$ step. The probability density for this case is plotted in Fig.~\ref{fig:rfrho1} at five times during one rf drive cycle, $\tau/\tau_p=0$, 1/4, 1/2, 3/4, and 1, with $\rho(\tilde Q)$ being identical at the beginning and end of the drive cycle. First one notes that the corresponding $\tilde Q(\tau)$ curve plotted in Fig.~\ref{fig:qt2}(d) shows regions where the quasicharge repeatedly lingers for an extended period near both $\tilde Q=-e/2$ and near $e/2$. This behavior explains why the probability distribution includes two peaks typically near these values of quasicharge. We can also see from Fig.~\ref{fig:qt2}(d) that single-electron tunneling often leads to repetitions of the plateau near $e/2$ but not the plateau near $-e/2$, and this explains why the probability peak near $e/2$ is larger. However, the distribution for $\tau=\tau_p/4$ violates these expectations. This anomaly is explained, however, when we consider the effect of the junction current, which shifts the entire distribution at a rate proportional to $i_j$. Between $\tau=0$ and $\tau_p/2$, the rf bias adds to the dc bias, and $i_j$ reaches a peak of about $i_0+i_1=0.65$ near $\tau=\tau_p/4$. As a result, both probability peaks are shifted by roughly $+e$ during this half cycle, with the larger peak turning into the smaller peak in the process. On the second half cycle, by contrast, the rf bias is negative, largely cancelling the positive dc bias, so the probability distribution is basically not shifted in quasicharge between $\tau=\tau_p/2$ and $\tau_p$. Thus, the changes in $\rho(\tilde Q)$ that occur over a drive cycle make sense in terms of the $\tilde Q(\tau)$ behavior shown in Fig.~\ref{fig:qt2}(d) and the shifts in quasicharge imparted by $i_j$.

The behaviour of the probability distribution for bias point (b) on the $n/m=1/1$ step, plotted in Fig.~\ref{fig:rfrho2}, is comparatively easy to understand. Here we find a single broad peak in $\rho(\tilde Q)$ that gradually shifts in quasicharge by $2e$ over the course of one drive cycle. Because $dq/d\tau=i_j\simeq i_0+i_1\sin(\omega\tau)$, we again expect this shift to be divided into a larger fraction that occurs during the first half cycle and a lesser fraction during the second half cycle, as seen in Fig.~\ref{fig:rfrho2}.

In the remainder of this paper, ensemble simulations based on relaxation to a periodic probability density become our primary tool for investigating Bloch steps in the rf-biased junction.

\section{Parameter Space}

Within the model considered here, the junction voltage is a function of the seven parameters listed in Eqs.~(\ref{eq:ej})--(\ref{eq:omega}). To gain a broader perspective on the nature and range of Bloch steps, we now examine voltage--current curves for a range of rf amplitudes and frequencies, $i_1$ and $\omega$, and normalized Josephson coupling energies $\varepsilon_j$, while setting the remaining parameters, $g_s$, $t_j$, and $\alpha$, to zero. Assuming the latter parameters are zero allows us to examine harmonic Bloch steps under ideal conditions. In particular, $g_s=0$ implies that the junction is perfectly isolated from its electromagnetic environment, $t_j=0$  eliminates thermally activated single-electron tunneling, and $\alpha=0$ eliminates Zener tunneling, forcing the junction to remain in the first energy band. However, all of the processes omitted here will be important in the following section, where we attempt to model experimental voltage--current curves.

\begin{figure*}[t]
\includegraphics{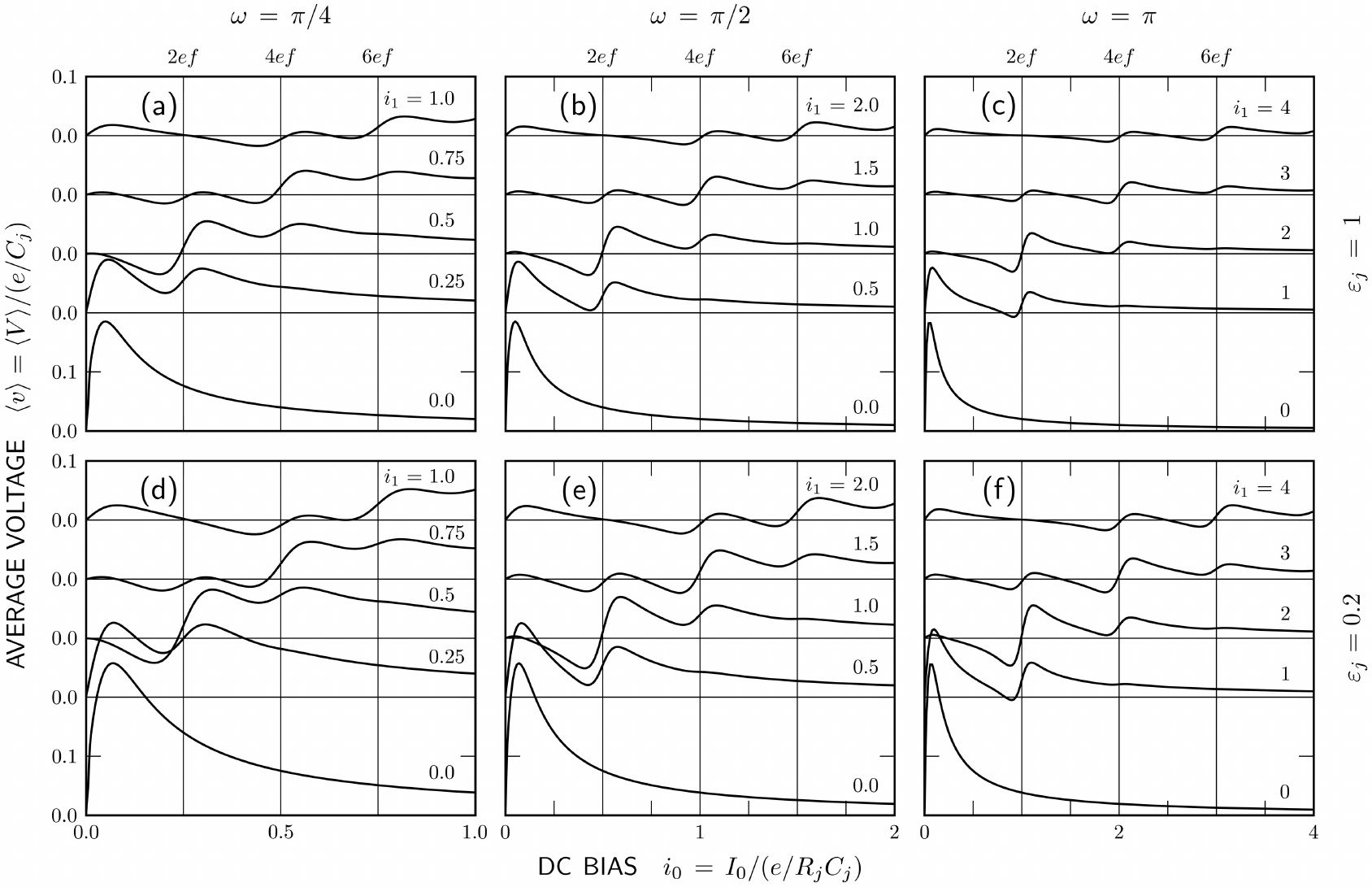}
\caption{\label{fig:ivarray} Average voltage as a function of dc bias in the case of an ideal current source, in the presence of single-electron tunneling and in the absence of Zener tunneling, for various values of Josephson coupling energy $\varepsilon_j$, rf freqency $\omega$, and rf amplitude $i_1$: ($\varepsilon_j$, $\omega$) = (a) (1, $\pi/4$); (b) (1, $\pi/2$); (c) (1, $\pi$); (d) (0.2, $\pi/4$); (e) (0.2, $\pi/2$); (f) (0.2, $\pi$). All curves derive from ensemble calculations with $n_q=100$ and $t_j=\alpha=g_s=0$. Vertical lines indicate the location of the first three harmonic Bloch steps at $I_0=2ef$, $4ef$, and $6ef$.}
\end{figure*}

Each of the six frames in Fig.~\ref{fig:ivarray} shows a collection of voltage--current curves corresponding to five different rf amplitudes. Because we have chosen $g_s=0$, none of the curves show subharmonic steps, but harmonic steps at bias currents $I_0=2nef$ for $n=1$, $2$, and $3$ are well represented, with higher-order steps appearing at higher rf amplitudes. As expected from our earlier example, however, all of these ``constant-current'' steps have a finite slope on the order of $R_j$ or less. This slope is in contrast to the constant-voltage steps of Josephson voltage standards in which deviations from the quantized voltage are experimentally undetectable over the central region of each step. Thus, while Bloch oscillations can be synchronized to some extent with an external rf bias, for the parameters considered here the resulting steps would not be useful as the basis of a precision current standard.

For the purpose of demonstrating the existence of Bloch oscillations, regardless of their utility as a current standard, Fig.~\ref{fig:ivarray} provides a guide to the selection of suitable parameters. For example, consider the drive frequency $\omega=2\pi fR_jC_j$. As Figs.~\ref{fig:ivarray}(a) and (d) suggest, if $\omega$ is too small then adjacent Bloch steps begin to overlap, so it is best to keep the separation between steps $2ef$ greater than the step width, which is on the of order the characteristic voltage $e/C_j$ divided by the slope $R_j$. When numerical factors on the order $1$ are eliminated, this condition reduces to $\omega\gtrsim1$. On the other hand, Figs.~\ref{fig:ivarray}(c) and (f) reveal that step amplitudes generally decrease with increasing frequency, so $\omega$ should not be too large. A second factor also sets an upper limit on $\omega$, namely the condition $I_j\ll e/R_KC_j$, required to insure that the system is always in a quasicharge eigenstate. Given that the current on the $n$th harmonic step is $I_j=2nef$, this condition reduces to $\omega\ll\pi R_j/nR_K$. Thus, the largest step amplitudes are expected to result for $\omega$ somewhat larger than $1$ but not too large.

Figure~\ref{fig:ivarray} does not, however, provide significant clues about what ratio $\varepsilon_j=E_j/E_c$ of Josephson to charging energy might optimise the amplitude of Bloch steps. The steps for $\varepsilon_j=0.2$ and $1$ shown here are not dramatically different. Instead, the optimum $\varepsilon_j$ is suggested by other constraints. In particular, our analysis is predicated on the condition that $\varepsilon_j\lesssim1$ in order that charge rather than phase be the dominant quantum variable. While it is not possible to specify a particular $\varepsilon_j$ beyond which the analysis breaks down, this parameter clearly should not be very much larger than $1$. At the same time, the probability of Zener tunneling between the first and second band is proportional to $\exp(-\varepsilon_j^2)$, so it is advantageous to make $\varepsilon_j$ as large as possible to suppress this unwanted process. Thus, the optimum $\varepsilon_j$ is on the order of $1$ and neither very much smaller nor very much larger than $1$.

Given that nonzero values of $g_s$, $t_j$, and $\alpha$ compromise the existence of Bloch steps and that $\omega$ and $\varepsilon_j$ are near their optimum values in Fig.~\ref{fig:ivarray}, we conclude that the steps shown here are typical of the strongest Bloch steps that can possibly be observed in a nanoscale Josephson junction.

\section{Experimental Comparison}

\begin{figure*}[t]
\includegraphics{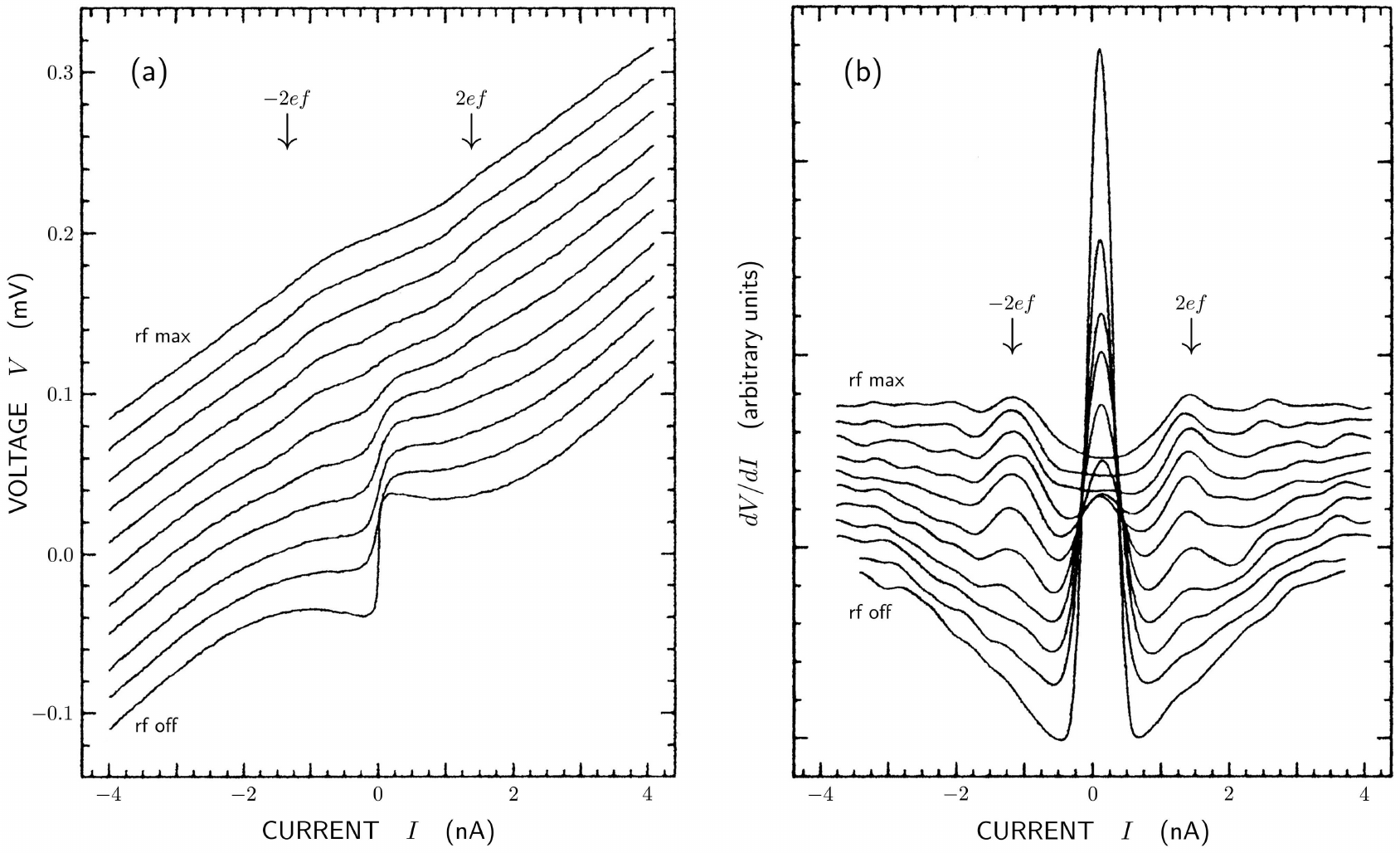}
\caption{\label{fig:expt} Experimental voltage--current curves (a) and their derivatives (b) at various rf power levels recorded by Kuzmin {\it et al.} for the nanoscale Josephson junction N1.\cite{kuz94b} The junction was nominally cooled to 60~mK and driven by 4~GHz rf power.  (Reproduced with the permission of Elsevier.)}
\end{figure*}

Having explored the inner workings of the model of nanoscale Josephson junctions introduced by GS,\cite{gei88} we now apply this model to the experimental results of Kuzmin {\it et al.}\cite{kuz94b} In particular, we consider the experimental results for the junction N1 shown in their Fig.~3 and reproduced here in Fig.~\ref{fig:expt}. Junction N1 is an Al/AlO$_x$/AlPbAu tunnel junction of area 0.01 $\mu$m$^2$ and is isolated from the surrounding electromagnetic environment by thin-film Cr resistors that are 0.1~$\mu$m by 6~nm in cross section and 10~$\mu$m in length. As shown in Fig.~\ref{fig:expt}, when cooled to a nominal temperature of 60~mK and driven by 4-GHz microwaves, this junction revealed strong evidence of Bloch steps at $I=2ef=1.28$~nA that is especially clear in the $dV/dI$ curves of Fig.~\ref{fig:expt}(b).

The question to be considered now is the extent to which the GS model might explain the experimental results of Kuzmin {\it et al.} Because the experimental parameters are not known with certainty, however, we attempt to make only a semi-quantitative comparison between theory and experiment, and simulation parameters were chosen to be representative rather than to produce a detailed fit to experiment. As Kuzmin {\it et al.} discuss, one experimental uncertainty is the junction temperature $T_j$. While the base temperature of their refrigerator is 60~mK, the power dissipated in the isolation resistors in the presence of dc and rf biases is likely to raise $T_j$ well above 60~mK. In particular, they estimate that the junction temperature may be as high as 300~mK for a bias current of 1.3~nA. In the following we assume that $T_j$ is independent of bias, which in itself precludes the possibility of a detailed fit to experiment.

Parameters of the experiment and simulation are listed in Table~\ref{tab:param}. On the experimental side, the junction capacitance $C_j$ was estimated from the junction area, and the energy-gap voltage $V_g$, Bloch-nose voltage $V_b$, and normal-state resistance $R_n$ were read from the voltage--current characteristic. The critical current was then estimated from the BCS relation $I_c=\pi V_g/4R_n$. Because Bloch steps occur at voltages less than 1~mV and  much less than $V_g$, the junction resistance $R_j$ relevant to Bloch oscillation is the sub-gap resistance. Typically the sub-gap resistance is much greater than $R_n$, but Zener tunnelling in this low-voltage region prevents reading $R_j$ from the voltage--current characteristic, and this parameter eludes experimental evaluation.

\begin{table*}
\caption{\label{tab:param}The experimental parameters for sample N1 of  Kuzmin {\it et al.}\cite{kuz94b} and the corresponding parameters adopted in our simulation. Formulas in parentheses to the right of numerical entries indicate the method of evaluation.}
\begin{ruledtabular}
\begin{tabular}{lr|cr|cr}
Parameter&&\multicolumn{2}{c|}{Experiment$\hspace{.4in}$}
&\multicolumn{2}{c}{Simulation$\hspace{.4in}$}\\ \hline
Junction area&$\mu$m$^2\hspace{.2in}$&0.01&&---\\
Junction capacitance, $C_j$&fF$\hspace{.2in}$&0.5&&0.5\\
Charging Energy, $E_c=e^2/2C_j$&$\mu$eV$\hspace{.2in}$&160&&160\\
Energy-gap voltage, $V_g=(\Delta_a+\Delta_b)/e$&$\mu$V$\hspace{.2in}$
&450&&---\\
Normal-state resistance, $R_n$&k$\Omega\hspace{.2in}$&7&&---\\
Junction critical current, $I_c$&nA$\hspace{.2in}$&50&($\pi V_g/4R_n$)$\hspace{.2in}$
&39&($2e\varepsilon_jE_c/\hbar$)\\
Josephson coupling energy, $E_j$&$\mu$eV$\hspace{.2in}$&100
&($\hbar I_c/2e$)$\hspace{.2in}$&80&($\varepsilon_jE_c$)\\
Bloch-nose voltage, $V_b$&$\mu$V$\hspace{.2in}$&40&&43&($v_be/C_j$)\\
Subgap junction resistance, $R_j$&k$\Omega\hspace{.2in}$&---&&100
&($\omega/2\pi fC_j$)\\
Source resistance, $R_s$&k$\Omega\hspace{.2in}$&130&&$\infty$&($R_j/g_s$)\\
Drive frequency, $f$&GHz$\hspace{.2in}$&4&&4\\
Junction temperature, $T_j$&mK$\hspace{.2in}$&60--300&&560&($t_jE_c/k$)\\ \hline
$\varepsilon_j=E_j/E_c$&&0.6&&0.5\\
$t_j=kT_j/E_c$&&0.03--0.16&&0.3\\
$\alpha=R_K/\pi^2R_j$&&---&&0.05\\
$g_s=R_j/R_s$&&---&&0\\
$\omega=2\pi R_jC_jf$&&---&&2$\pi$/5\\
\end{tabular}
\end{ruledtabular}
\end{table*}

\subsection{Model Parameters}

\begin{figure}[b]
\includegraphics{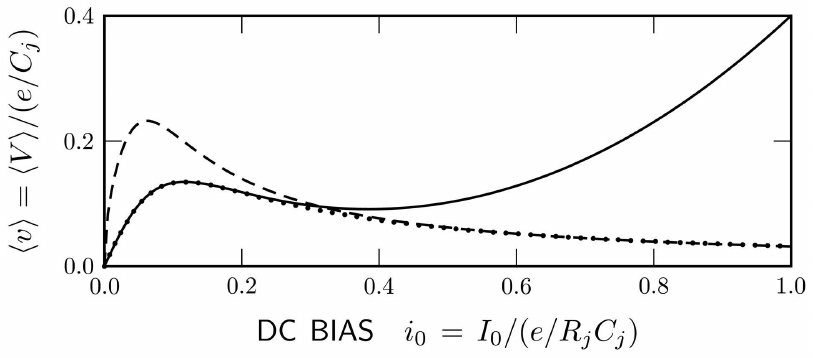}
\caption{\label{fig:dciv3} Voltage--current curves computed by ensemble simulation for $\varepsilon_j=0.5$, $g_s=0$, and variously with $t_j=\alpha=0$ (dashed curve), $t_j=0.3$ and $\alpha=0$ (dotted curve), and $t_j=0.3$ and $\alpha=0.05$ (solid curve). All computations are for $n_b=5$ and $n_q=100$.}
\end{figure}

The five dimensionless parameters that enter into the GS model are listed at the bottom of Table~\ref{tab:param}. Because the experimental Bloch steps show no sign of the sharp structures associated with a finite source conductance, we have chosen to set $g_s=0$. Also, while Kuzmin {\it et al.} estimate $\varepsilon_j$ at 0.6, we have arbitrarily stepped it down to 0.5. The remaining three parameters were chosen as follows, based on matching the experimental dc voltage--current characteristic.  The simulated dc $\langle v\rangle$--$i_0$ curve for $\varepsilon_j=0.5$ and $t_j=\alpha=0$ is shown by the dashed line in Fig.~\ref{fig:dciv3}. Using the experimental value of $C_j$, the Bloch-nose voltage for this curve is $V_b=ev_b/C_j=74~\mu$V, or almost twice the experimental value. This voltage can be adjusted downward by increasing the temperature,  and by trial and error we find that $t_j=0.3$ reduces $V_b$ to 43~$\mu$V, as indicated by the dotted curve in Fig.~\ref{fig:dciv3}. Similarly, by increasing the Zener tunneling parameter $\alpha$ from 0 to 0.05, we can create a broad minimum in the $\langle v\rangle$--$i_0$ characteristic (solid curve in Fig.~\ref{fig:dciv3}) that mimics the minimum in the experimental characteristic of Fig.~\ref{fig:expt}(a). Finally, noting that the first Bloch step occurs experimentally near this voltage minimum, we choose the dimensionless frequency parameter $\omega$ to place the first step at $i_0=0.4$ in Fig.~\ref{fig:dciv3}, so that $\omega=\pi i_0=2\pi/5$.

The dimensioned parameters implied by our chosen set of dimensionless parameters are also listed in Table~\ref{tab:param}. Here we have adopted the experimental values for $C_j$, $E_c$, and $f$ and combined them with the dimensionless parameters of the model according to the formulas in parentheses to fill in the remaining dimensioned quantities. In general, these derived quantities are in reasonable agreement with experimental values. However, the junction temperature of 560~mK assumed in the simulation is almost twice the experimentally estimated temperature on the first Bloch step. Also, the shunt resistor $R_s$ of our Norton equivalent drive circuit is nominally identical to the series isolation resistance of the experimental circuit. By choosing $g_s=0$, we have made this isolation resistance infinite, but the consequences should be minimal.  We additionally note that, while the Zener tunneling constant $\alpha$ has been taken as a free parameter here, it is actually defined as a simple function of $R_j$. Inverting this equation to solve for $R_j$, we find that $\alpha=0.05$ corresponds to $R_j=52$~k$\Omega$, or about half the value indicated in Table~\ref{tab:param}.

One disturbing feature of the simulated dc $\langle v\rangle$--$i_0$ characteristic (solid curve in Fig.~\ref{fig:dciv3}) in comparison with the corresponding experimental curve of Fig.~\ref{fig:expt}(a) is the relatively slow initial rise of the simulated curve. This discrepancy is probably explained by two assumptions that we have made in applying the GS model. First, by assuming a fixed temperature, we ignore the fact that near zero bias there is no significant dissipation in the isolation resistors, so the junction temperature here will be closer to 60~mK than 560~mK. If this lower temperature were taken into account, the initial slope would be closer to that of the steeper dashed curve in Fig.~\ref{fig:dciv3} for $t_j=0$ than the solid curve for $t_j=0.3$. Second, by assuming $g_s=0$, we have eliminated the possibility of an initial spike near $i_0=0$, like that shown in Fig.~\ref{fig:dciv2}, which might also contribute to the rapid initial rise of the experimental curve. Nevertheless, with the assumptions and parameters chosen for our GS model, we expect that an rf bias will evoke simulated Bloch steps similar to those observed experimentally.

\subsection{Simulated Bloch Steps}

\begin{figure*}[t]
\includegraphics{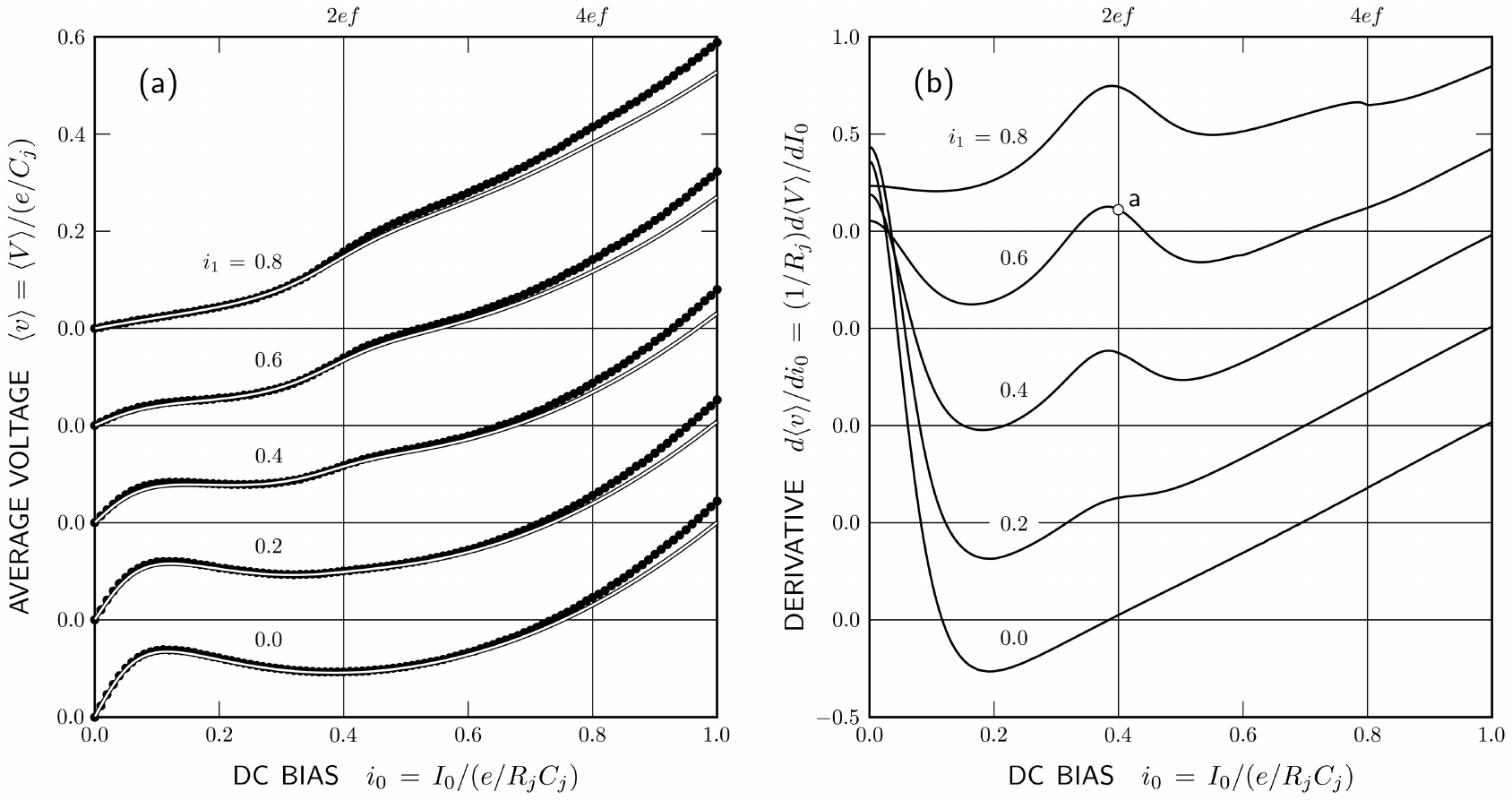}
\caption{\label{fig:ivsim} Simulated voltage--current curves (a) and their derivatives (b) for various rf amplitudes. Results are shown for $\varepsilon_j=0.5$, $t_j=0.3$, $\alpha=0.05$, $g_s=0$, and $\omega=2\pi/5$. In (a), black dots are for Monte Carlo simulations with an averaging time of $10^6$ drive cycles and white lines are for ensemble calculations with $n_b=5$ and $n_q=100$. Frame (b) shows the result of numerically differentiating the ensemble curves in (a). Vertical lines mark the expected locations, $I_0=2ef$ and $4ef$, of the first and second Bloch steps.}
\end{figure*}

This expectation is largely met by the simulated voltage--current curves and corresponding derivative curves shown in Fig.~\ref{fig:ivsim} for several microwave drive amplitudes. Frame (a) shows $\langle v\rangle$--$i_0$ curves for both Monte Carlo and ensemble calculations, indicated by black dots and white lines respectively. In these curves, the first microwave induced Bloch step is subtly apparent as a slight increase in slope near $I_0=2ef$ or $i_0=\omega/\pi=0.4$, the expected location. Although the Monte Carlo and ensemble results diverge somewhat at higher dc bias (probably due to the relatively small number of quasicharge bins used in the ensemble calculations), they closely agree with regard to the general appearance of the first step. Moreover, both calculations are in qualitative agreement with the experimental results shown in Fig.~\ref{fig:expt}(a), where the first step is also apparent only on close inspection.

On the other hand, the derivative curves, $dV/dI$ and $d\langle v\rangle/di_0$, shown for the experiment in Fig.~\ref{fig:expt}(b) and for the ensemble simulation in Fig.~\ref{fig:ivsim}(b), give dramatic evidence for Bloch oscillations near the expected dc bias and are in excellent qualitative agreement with one another. In particular, we note that the simulated derivative curve for $i_0=0.8$ is much like the three experimental curves at the highest rf power levels. In this case, experimental and simulated curves show three points of strong agreement. First, the peak at $I=0$, prominent at lower rf power, is almost entirely suppressed. Second, the peak at $I=2ef$ is near its maximum amplitude. Third, the width of the first peak is roughly $\Delta I=0.4ef$ for both the experimental and simulated curves. Similar agreement, is found at lower rf power between the simulated derivative for $i_1=0.4$ and the middle experimental curve (the 6th curve counted either up from ``rf off '' or down from ``rf max''). Here the peak at $I=0$ and that at $I=2ef$ are both well developed, with the amplitude of the former being 2 to 3 times that of the latter. At yet lower rf amplitudes, however, the simulated derivative at $I=0$ is much less than that observed experimentally. As discussed previously, this discrepancy may result because the simulations assume a constant junction temperature, while the experimental temperature probably falls rapidly as the rf and dc levels approach zero.

Another point of qualitative agreement between the simulated and experimental $dV/dI$ curves is a tendency for the peak associated with the first step to occur at a dc bias that is 3 to 4\% below $2ef$. This effect is seen consistently in simulations and also appears in several experimental $dV/dI$ curves.

While we have so far claimed only qualitative agreement between simulation and experiment, it is important to note that the scale factors that convert the dimensionless voltage $v$ and dc bias $i_0$ to real voltages and currents are $e/C_j=0.32$~mV and $e/R_jC_j=3.2$~nA. When these are applied to the simulated $\langle v\rangle$--$i_0$ curves of Fig.~\ref{fig:ivsim}(a), one finds that the ranges of voltage and current being plotted are comparable to the experimental curves of Fig.~\ref{fig:expt}(a). Thus, the agreement between simulation and experiment is in fact semi-quantitative. Although closer quantitative agreement would require treating the dependence of junction temperature on the power dissipated in the isolation resistors, there seems little doubt that the GS model implemented here explains the basic features of the experimentally observed steps and in doing so confirms and strengthens their interpretation in terms of Bloch oscillations.

\subsection{Zener and Thermally Assisted Tunneling}

As a final note on our simulations, we examine in further detail two processes, Zener tunneling and thermally assisted single-electron tunneling, that were introduced in Fig.~\ref{fig:ivsim} by nonzero values for $\alpha$ and $t_j$ and have not been explored in previous cases. These processes allow the system to access bands above the first energy band and are important to the overall agreement between the simulated and experimental voltage--current characteristics.

\begin{figure}[t]
\includegraphics{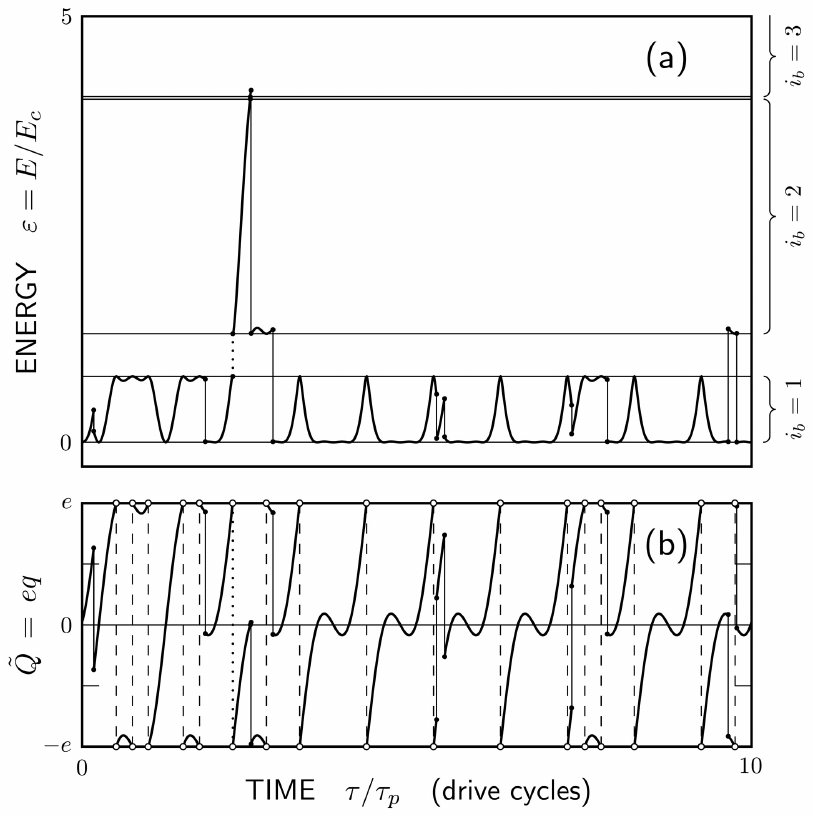}
\caption{\label{fig:qt4}Monte Carlo results for the energy (a) and quasicharge (b) as a function of time for bias point {\bf a} in Fig.~\ref{fig:ivsim}(b), that is for $\varepsilon_j=0.5$, $t_j=0.3$, $\alpha=0.05$, $g_s=0$, $i_0=0.4$, $i_1=0.6$, and $\omega=2\pi/5$. Solid vertical lines indicate single-electron tunneling, dashed vertical lines indicate Bloch reflections,  and dotted vertical lines indicate Zener tunneling. Horizontal lines mark the limits of the first three energy bands in frame (a).}
\end{figure}

A brief examination of Zener and thermally activated tunneling is given in Fig.~\ref{fig:qt4}, which plots the time evolution of the energy and quasicharge generated by Monte Carlo simulation. The plot displays behavior characteristic of the bias point on the first Bloch step labeled {\bf a} in Fig.~\ref{fig:ivsim}(b). At this bias point, the ensemble calculation reveals that on average the junction spends 98.44\% of its time in the first energy band, 1.53\% in the second band, and 0.03\% in the third. During the ten drive cycles shown in Fig.~\ref{fig:qt4}, we see from the energy plot that the junction leaves the first band only twice: for brief intervals during the third and tenth drive cycles. As expected for a bias point on the first Bloch step, however, the junction spends most of its time experiencing a single Bloch oscillation during each drive cycle, as in cycles 5 through 7. Here, the dc and rf biases combine during the first half of the drive cycle to push the quasicharge to $e$, where it Bloch reflects to $-e$, while during the second half cycle the dc and rf biases largely cancel and the quasicharge merely oscillates around 0. Occasionally, however, this process is flipped and the quasicharge oscillates around $\pm e$ during the second half of the cycle, as during drive cycles 1 and 2.

The event of primary interest in Fig.~\ref{fig:qt4}, is the Zener tunneling during drive cycle 3. In this case, when the combined dc and rf biases push the quasicharge to $e$, the junction Zener tunnels to the second band. The probability of Zener tunneling between bands 1 and 2 is $P_Z=0.29$ when $i_j$ is near its maximum, so tunneling to the second band is not uncommon. After Zener tunneling, however, the bias current pushes the junction to the top of the second band and it Zener tunnels for a second time to the third band. Because the gap between the second and third bands is small, the tunneling probability is $P_Z= 0.94$, and tunneling here is likely. The junction remains in the third band only briefly, however, before single-electron tunneling returns it to the bottom of the second band, and a short time later a second single-electron tunneling returns the junction to the bottom of the first band. This kind of brief excursion into higher bands is relatively frequent at this bias point, even though the junction spends most of its time in the first band.

In contrast, the interband transition that occurs during the tenth drive cycle is highly unusual. Here, the junction jumps from the bottom of the first band to the bottom of the second band by thermally activated single-electron tunneling. Because the change in energy  $\Delta\varepsilon=1.32$ for this jump is much larger than the thermal energy $t_j=0.3$, the tunneling rate is very small. Nonetheless, such rare events are bound to occur from time to time.

\section{Conclusion}

To summarize, we have calculated voltage--current characteristics of nanoscale Josephson junctions whose charging energy is greater than or comparable to the Josephson energy using two separate approaches: Monte Carlo and ensemble calculations. While Monte Carlo calculations follow the dynamics of the quasicharge state of a single junction in time, the ensemble approach looks at the distribution of quasicharge states within an ensemble of junctions as function of time. Although the two approaches are equivalent in principle, each has its own computational advantages, and numerical results sometimes differ slightly. By including the shot noise of quasiparticle tunneling and the possibility of Zener tunneling to higher bands, these calculations demonstrate that the originally proposed \cite{lik84,ave85,lik85,ave86} Bloch steps of fixed slope are destroyed by these error processes. Using the ensemble approach, we are able to create a parameter map of the voltage--current characteristics and show how the height and the width of the Bloch steps vary with different junction parameters such as the ratio of Josephson to charging energy, applied microwave power and frequency. However, the Monte Carlo approach allows us to follow the evolution of a junction's quasicharge in time under the influence of microwaves and understand the mechanisms of phase locking. Based on this analysis, we can explain the harmonic and subharmonic steps that occur with a finite  source resistance. In the end we show that our calculations semi-quantitatively explain the experimental results of Kuzmin {\it et al.}\cite{kuz94b}

One of the important conclusions drawn from our calculations is that, for a fixed sub-gap conductance $G_j$, even in the limiting case of zero temperature and the absence of Zener tunneling, where the junction state is confined to the lowest energy band, quasiparticle tunneling can still broaden the Bloch steps to such an extent that it renders them unusable for a precise metrological current standard. Our quasicharge versus time plots clearly show that single-electron tunneling is the primary source of disruption to the locking behavior required for Bloch steps. Moreover, as revealed by bias points (d) and (e) of Fig.~\ref{fig:qt2}, single-electron tunneling can be problematic for both subharmonic and harmonic steps, regardless of how steep they appear in the voltage--current characteristic. Roughly speaking, a current standard with a precision of say a part in $10^6$ would correspond to one quasiparticle tunneling event per $10^6$ drive cycles. According to the BCS theory, however, the density of quasiparticles is expected to decay exponentially in the limit of low temperature, $n_{qp}\propto\exp[-(\Delta_a+\Delta_b)/2kT_j)]$. In this case $G_j$ would vanish at typical dilution refrigerator temperatures ($<$100\,millikelvin), corresponding to single-electron tunneling errors to levels that might permit metrology. 

It must be noted, however, that the question of reducing quasiparticle densities requires careful device engineering. While thermal quasiparticles can, in principle, be eliminated by cooling to dilution refrigerator temperatures, Joule heating in on-chip bias resistors will always provide a local source of heat. As Kuzmin {\it et al.}\cite{kuz94a} originally noted, the series isolation resistors in their experiment posed a significant problem in achieving low enough temperatures to exclude the presence of thermal quasiparticles. Meanwhile there is overwhelming evidence that a significant density of nonequilibrium quasiparticles is universally present in ultra-small Josephson junction devices such as qubits \cite{palmer2007steady,sha08,mar09,catelani2011quasiparticle,sun2012measurements,voo14} and Cooper pair transistors \cite{aumentado2004nonequilibrium,sillanpaa2004inductive,ferguson2006microsecond}, although improvements in filtering and shielding from stray infrared radiation have improved this situation considerably. While it is conceivable that nonequilibrium quasiparticle tunneling rates can be reduced to metrological levels, the problem of Joule heating and thermal quasiparticle generation will remain in any scheme that seeks to generate nanoampere-level currents sourced through on-chip bias resistors. Other experiments\cite{gui10,mas12,wei15} have attempted to avoid the use of such resistors by constructing so-called ``superinductances'' to create a high-impedance environment while minimizing power dissipation. In our view, a successful strategy will, at the outset, identify the need to mitigate the inevitable problem that quasiparticles pose.


\begin{acknowledgments}
We are pleased to thank Leonid Kuzmin for granting permission to republish the experimental data appearing in Fig.~\ref{fig:expt}. This work was supported by NIST-on-a- Chip Initiative. This article is a contribution of the U.S. government, not subject to U.S. copyright.
\end{acknowledgments}

\bibliography{bloch}

\end{document}